\documentclass[prd,preprintnumbers,showpacs,amsmath]{revtex4}
\usepackage[dvips]{epsfig}

\begin{document}

\preprint{  CERN-PH-TH/2005-204}

\title{Gauge invariance properties and singularity cancellations in a modified PQCD}

\author{Alejandro Cabo$^{1,2}$ and Marcos Rigol$^{3}$ \medskip}

\affiliation{$^{1}$Theory Division, CERN, Geneva, Switzerland}

\affiliation{$^{2}$ Group of Theoretical Physics, Instituto de
Cibern\'{e}tica, Matem\'{a}tica y F\'{\i}sica,\\ Calle E, No. 309,
Vedado, la Habana, Cuba}

\affiliation{$^{3}$ Physics Department, University of California,
Davis, CA 95616, USA}

\begin{abstract}
\noindent The gauge-invariance properties and singularity elimination of the
modified perturbation theory for QCD introduced in previous works, are
investigated. The construction of the modified free propagators is
generalized to include the dependence on the gauge parameter $\alpha $.
Further, a functional proof of the independence of the theory under the
changes of the quantum and classical gauges is given. The singularities
appearing in the perturbative expansion are eliminated by properly combining
dimensional regularization with the Nakanishi infrared regularization for
the invariant functions in the operator quantization of the $\alpha $%
-dependent gauge theory. First-order evaluations of various quantities are
presented, illustrating the gauge invariance-properties.
\end{abstract}

\pacs{12.38.Aw,12.38.Bx,12.38.Cy,14.65.Ha}

\maketitle

\section{Introduction}

Improving the understanding of chromodynamics is an important issue in
modern Particle Physics \cite{Shuryak}. Although the small couplings at high
momenta allow the use of perturbation theory in this region, the expansion
is however unable to describe the physics at low energies. \ In former works
\cite{1995,PRD,epjc,tesis,jhep} we have been considering the development of
a modified perturbation expansion, searching to describe at least some
low-energy properties. The proposed scheme \cite{1995} conserves colour and
Lorentz invariances. Indeed, it was motivated by the aim of eliminating the
symmetry limitations of the earlier chromomagnetic field models \cite
{Savvidy1,Savvidy2,Savvidy3,Reuter}. The Feynman expansion proposed in Ref.
\cite{1995} included in the gluon propagator a term produced by the
condensation of zero-momentum gluons. Later, in Refs. \cite{Hoyer1,jhep} the
interest of also including quark condensates was argued. Just in the first
approximation, the modified rules produced a non-vanishing gluon
condensation parameter $\langle g^2G^2\rangle $ \cite{Zakharov}. Afterwards,
in Ref. \cite{1995}, a non-zero tachyon mass for the gluons was evaluated at
the one-loop level. This outcome was consistent with the conclusion in Ref.
\cite{fukuda} about the tachyon character of the normal Green functions in
the presence of a gluon condensate \cite{tachyon1,tachyon2}. Also, the
one-loop effective potential as a function of the condensate parameter
indicated its dynamic generation. The plausibility of modifying the Feynman
rules of QCD, including gluon condensation, was also foreseen in Ref. \cite
{Hoyer}. In this work a filling of gluon states for all momenta $k<\Lambda _{%
\text{qcd}}$, was assumed in defining an alternative free vacuum state, upon
which to connect the colour interaction.

The addition of \ a delta-function term at zero momentum in the gluon
propagator had also been investigated in Refs. \cite{Munczek,Burden}. The
aim was to develop a \ phenomenological model \ for meson resonances,
showing a confining inter-quark potential. \ \ The multiplicative constant
of the delta function had the opposite sign with respect to the one chosen
in Refs. \cite{PRD,tesis}. For this selection of the sign, the singularities
of the quark propagator showed no pole on the real $\ p^2$ axis \cite{Burden}%
. Thus, the results led to a physical description different from that in
Refs. \cite{1995}-\cite{tesis}.

The expansion examined in Ref. \cite{1995} was derived in Refs. \cite
{PRD,tesis} as the result of the adiabatic connection of the colour
interaction, upon a particularly constructed free physical state. The \
operator quantization \ scheme of Kugo and Ojima \cite{Kugo,OjimaTex}
allowed this state to be determined by the creation of zero-momenta gluons
and ghost particles. \ The propagators considered in Ref. \cite{1995}, then
emerged as generating the Wick expansion. The annihilation of the modified
vacuum by the action of the Becchi-Rouet-Stora-Tyutin (BRST) charge assured
its physical character at zero coupling. It also followed that the \
parameter describing the condensation should be real and positive. This
specific definition, in our case, discarded the possibility of fixing the
sign of the factor multiplying the delta function, in the way chosen in \
Refs.\ \cite{Munczek,Burden}.{\ }\ It should be added that the propagators
employed by Munczek and Nemirovsky were also obtained in Ref. \cite{pavel},
as generated by the Wick expansion from a vacuum including condensed gluons
in a squeezed state. It may be that the BRST physical condition also could
be satisfied by that class of states. Then, their alternative approach could
be constructed along the same lines. The difference in the end results for
these two treatments might be related with the fact that two different kinds
of `squeezed' gluon states were employed \cite{pavel}.

In the present work, we consider the investigation of the gauge-invariance
and \ regularization properties of the modified expansion under study. Our
main objectives will be: \ a) to generalize the procedure to arbitrary
values of the gauge parameter $\alpha $, \ b) \ to eliminate the infrared
singularities appearing as a consequence of the delta-function structure of
the new propagators, c) to present the formal proof of the \
Ward-Takahashi-Slavnov identities and the gauge-parameter independence of
the physical quantities, and \ finally d) to check the satisfaction of the
gauge independence \ of the effective action and the transversality of the
gluon self-energy at zero values of the mean fields, to second order in the
coupling $g$ and all orders in the expansion in the condensate parameters.

Section 2 starts by presenting the initial state to be used in the Wick
expansion formula. It is then employed in the derivation of the modified
propagators. At this point, the \ invariant functions appearing as the
results of the commutators defining all the propagators are modified by
implementing the Nakanishi procedure for the infrared regularization of
these functions \cite{Nakanishi}. \ This procedure allows us to show that
all the Feynman parts of the propagators \ are regularized to be equal to
zero in a small neighbourhood of the origin in momentum space. The size of
the vanishing region is fixed by the Nakanishi regularization parameter $%
\sigma $. \ This is a basic outcome that \ further allows the elimination of
a large class of the singularities generated by the delta-function character
of the modified propagators.

Section 3 is devoted to deriving the Ward-Takahashi-Slavnov (WTS) identities
corresponding to the Green functions in the modified expansion \cite
{boulware, dewitt,thooft}. They fully coincide with those in the usual
theory (PQCD). This outcome is suggested after considering that the modified
propagators are allowed inverse kernels of the tree-level inverse
propagators of massless QCD. \ The generating functional of the Green
functions is transformed into a Feynman integral form, only differing in the
boundary conditions on the fields, from the standard one in PQCD.

In Section 4, we formulate a procedure for the elimination of the infrared
singularities appearing in any order of the perturbative expansion due to
the $\delta (p)$ form of the condensate terms. The properties allowing the
cancellation of those divergences are two. The first is the use of
dimensional regularization for the Dirac's delta function evaluated at zero
momentum. As was shown in Ref. \cite{liebbrandt}, in the case of spatial
arguments, these expressions can be made to vanish in dimensional
regularization. An identical proof is employed \ here to argue the same
conclusion for momentum-space arguments. Using this rule, a large class of
the singular terms \ are removed. The second recourse leading to a vanishing
result for the remaining singular terms is the Nakanishi infrared
regularization described in Section 2.\ Specifically, the equality to zero
of the Feynman contribution to the gluon, quark and ghost propagators in the
neighbourhood of $p=0$ \ (of a size fixed by the Nakanishi parameter $\sigma
$) allows us to cancel the remaining singularities. Since these singular
terms arise from the evaluation of the Feynman propagators at zero momenta,
this is simply obtained after setting the absolute values of \ the momenta
at which the condensate states are created, being sufficiently smaller than $%
\sigma $.

Then, Section 6 presents the checking of the transversality of the
polarization tensor and the gauge-parameter independence of the effective
action evaluated at zero value of the mean fields. This is done here up to
second order in the coupling constant and any order in the condensate
parameters. Afterwards, comments on the results and possible directions of
their future extension are given in a summary.

Finally a resume of the Kugo-Ojima quantization procedure for free massless
QCD for general value of the gauge parameter $\alpha ,$ and a complementary
calculation, are given in Appendices A and B, respectively.

\section{ Modified Feynman expansion}

\subsection{The initial vacuum for the Wick series}

Let us consider in what follows the selection of the initial
vacuum state of free QCD upon which the interaction will be
connected. It will be a generalization of the previously defined
one in Refs.\ \cite{PRD,tesis}, but  physically equivalent to it,
since its physical part (the transverse gluons part) is not
modified. The changes appear only in the non-physical sector,
which we choose the most general possible in order to have a wider
freedom in the selection of the resulting propagators. The main
elements of the Kugo$-$Ojima operator quantization of the free
massless QCD for an arbitrary gauge parameter $\alpha$ are
reviewed in Appendix A. The resume collects the equations of
motion and commutation relations between the fields and the
creation and annihilation operators. The definitions of the
invariant functions and the various sorts of wave packets to be
employed below can also be found there. The vacuum state will be
defined in the form
\begin{eqnarray}
&&\mid\Psi \rangle =\exp \left\{ \sum\limits_{a=1}^8\sum\limits_{\vec{p}
_i,\left| \vec{p}_i\right| =P<\sigma}\left[ \sum\limits_{\lambda =1,2}\frac{%
C_\lambda \left( P\right) }2A_{\vec{p}_i,\lambda }^{a+}A_{\vec{p}_i,\lambda
}^{a+}+C_3\left( \alpha ,P\right) \left( B_{\vec{p}_i}^{a+}A_{\vec{p}
_i}^{L,a+}+i\overline{c}_{\vec{p}_i}^{a+}c_{\vec{p}_i}^{a+}\right) \right.
\right.  \nonumber \\
&&\qquad\qquad\qquad\qquad \qquad\qquad \left. \left. +D\left( \alpha
,P\right) B_{\vec{p}_i}^{a+}B_{\vec{p} _i}^{a+}\right] +
\sum_{s=1,2}\sum\limits_{\vec{p}_{i},\vec{q}_{i},\left\vert \vec{p} _{i},%
\vec{q}_{i}\right\vert =P<\sigma}C_{\vec{p}_{i},\vec{q}_{i}} b_{\vec{q}%
_{i}}^{s+} a_{\vec{p}_{i}}^{s+}\right\} \mid 0\rangle .  \label{Vacuum1}
\end{eqnarray}
The gluon (first part of the exponential) and quark (second part
of the exponential) pairs defining the state are created for a set
of  small but non-vanishing  momenta $\vec{p}_{i}$ in order to
avoid the difficulties occurring if they are directly chosen to
have zero momentum. A sum over the set $\vec{p}_{i}$ for
$\left\vert \vec{p}_{i}\right\vert =P<\sigma$ is introduced, also
to  have freedom in eliminating any preferential direction in the
space, after the Nakanishi parameter will be taken in the limit
$\sigma\rightarrow0$.

In order to simplify the exposition, we evaluate gluon expressions
below for a generic mode of given colour $a$ and momentum $\vec
{p}_{i}$ indices. This can be done because for the free theory,
and for the specific quantities to be  evaluated, contributions of
different modes can be worked out independently, thanks to the
commutation relations (\ref{commu}). At the necessary point in the
discussion, all  contributions will be included.

The proof that the proposed state (\ref{Vacuum1}) satisfies the
required BRST physical-state conditions
\[
Q_B\mid \Psi \rangle =i\sum\limits_{\vec{k},a}\left( c_{\vec{k}}^{a+}B_{\vec{%
k}}^a-B_{\vec{k}}^{a+\ }c_{\vec{k}}^a\right) \mid \Psi \rangle =0,\qquad
Q_C\mid \Psi \rangle =i\sum\limits_{\vec{k},a}\left( \overline{c}_{\vec{k}%
}^{a+}c_{\vec{k}}^a+c_{\vec{k}}^{a+}\overline{c}_{\vec{k}}^a\right) \mid
\Psi \rangle =0,
\]
proceeds as follows.

Using the commutation relations (\ref{commu}), the state $\mid \Psi \rangle$
in Eq.\ (\ref{Vacuum1}) can be written in the form
\[
\mid \Psi \rangle =\exp \left( D\left( \alpha,P\right) B_{\vec{p}_i}^{a+}B_{%
\vec{p}_i}^{a+}\right) \mid \Phi \rangle,
\]
where $\mid \Phi \rangle$ has the form
\[
\mid \Phi \rangle =\exp \left[ C_1\left( P\right) A_{\vec{p}_i,1}^{a+}A_{%
\vec{p}_i,1}^{a+}+C_2\left( P\right) A_{\vec{p}_i,2}^{a+}A_{\vec{p}
_i,2}^{a+}+C_3\left( \alpha ,P\right) \left( B_{\vec{p}_i}^{a+}A_{\vec{p}
_i}^{L,a+}+i\overline{c}_{\vec{p}_i}^{a+}c_{\vec{p}_i}^{a+}\right) \right]
\mid 0\rangle,
\]
and satisfies the physical-state conditions, as was probed in
Refs. \cite{PRD,tesis}
\[
Q_B\mid \Phi \rangle =Q_C\mid \Phi \rangle =0.
\]

The commutation relations (\ref{commu}) also allow to show that
\[
\left[ Q_B,\exp \left( D\left( \alpha,P\right) B_{\vec{p}_i}^{a+}B_{\vec{p}
_i}^{a+}\right) \right] =0,\qquad \left[ Q_C,\exp \left( D\left(
\alpha,P\right) B_{\vec{p}_i}^{a+}B_{\vec{p} _i}^{a+}\right) \right] =0.
\]
Hence
\[
Q_B\mid \Psi \rangle =\exp \left( D\left( \alpha,P\right) B_{\vec{p}
_i}^{a+}B_{\vec{p}_i}^{a+}\right) Q_B\mid \Phi \rangle =0, \qquad Q_C\mid
\Psi \rangle =\exp \left( D\left( \alpha,P\right) B_{\vec{p} _i}^{a+}B_{\vec{%
p}_i}^{a+}\right) Q_C\mid \Phi \rangle =0.
\]
The state (\ref{Vacuum1}) is then a physical state of the free theory. The
non-physical sector of $\mid \Psi \rangle$ is undetectable in the physical
world, and can neglected when calculating physical quantities like the norm,
energy, and particle number.

As stated before the physical sector of Eq.\ (\ref{Vacuum1}) is the same
proposed in previous work \cite{PRD,tesis}, then for the calculation of the
norm a similar result is obtained
\[
N=\langle \Psi \mid \Psi \rangle =\prod\limits_{\lambda
=1,2}\prod\limits_{a=1}^8\prod\limits_{\vec{p}_i,\left| \vec{p}_i\right|
=P<\sigma }\left[ \sum\limits_{m=0}^\infty \left| C_\lambda \left( P\right)
\right| ^{2m}\frac{\left( 2m\right) !}{\left( m!\right) ^2}\right] ,\text{%
\quad for }\left| C_\lambda \left( P\right) \right| <1.
\]

The normalized physical vacuum state is defined as
\begin{equation}
\mid \tilde{\Psi}\rangle =\frac 1{\sqrt{N}}\mid \Psi \rangle.
\end{equation}

The modifications introduced in the usual perturbation theory by the state (%
\ref{Vacuum1}) are calculated in this section for arbitrary values
of the gauge parameter $\alpha $. For this purpose the expressions
derived in Refs.\ \cite{Gasiorowicz,tesis} for the Green functions
generating functional of a free theory are used {\setlength
\arraycolsep{0.7pt}
\begin{eqnarray}
Z_0[j,\xi ,\overline{\xi },\eta ,\overline{\eta }] &=&Z_0^g[j]\ Z_0^q[\eta ,%
\overline{\eta }]\ Z_0^{gh}[\xi ,\overline{\xi }],  \label{zt} \\
Z_0^g[j] &=&\langle \tilde{\Psi}|T\exp \left( i\int dx\ j^\mu \left(
x\right) A_\mu \left( x\right) \right) |\tilde{\Psi}\rangle =\langle \tilde{%
\Psi}|\exp \left( i\int dx\ j^\mu \left( x\right) A_\mu ^{-}\left( x\right)
\right) \exp \left( i\int dy\ j^\mu \left( x\right) A_\mu ^{+}\left(
x\right) \right) |\tilde{\Psi}\rangle   \nonumber \\
&&\times \exp \left( \int dx\ dy\ \theta (y_0-x_0)j^\mu \left( x\right)
\left[ A_\mu ^{-}\left( x\right) ,\,A_\nu ^{+}\left( y\right) \right] \
j^\nu (y)\right) , \\
&=&Z_0^{g,m}[j]Z_0^{g,F}[j],  \label{fey1} \\
Z_0^q[\eta ,\overline{\eta }] &=&\langle \tilde{\Psi}|\exp \left( i\int
dx\left[ \overline{\eta }\left( x\right) \psi ^{-}\left( x\right) +\overline{%
\psi }^{-}\left( x\right) \eta \left( x\right) \right] \right) \exp \left(
i\int dy\left[ \overline{\eta }\left( x\right) \psi ^{+}\left( x\right) +%
\overline{\psi }^{+}\left( x\right) \eta \left( x\right) \right] \right) |%
\tilde{\Psi}\rangle   \nonumber \\
&&\hspace{-1.5cm}\times \exp \left( \int dx\ dy\ \overline{\eta }(y)\left[
\theta (x_0-y_0)\left\{ \psi ^{-}\left( y\right) ,\overline{\psi }^{+}\left(
x\right) \right\} -\theta (y_0-x_0)\left\{ \psi ^{+}\left( y\right) ,%
\overline{\psi }^{-}\left( x\right) \right\} \right] \eta (x)\right) , \\
&=&Z_0^{q,m}[\eta ,\overline{\eta }]Z_0^{q,F}[\eta ,\overline{\eta }], \\
Z_0^{gh}[\xi ,\overline{\xi }] &=&\langle \tilde{\Psi}|\exp \left( i\int
dx\left[ \overline{\xi }\left( x\right) c^{-}\left( x\right) +\overline{c}%
^{-}\left( x\right) \xi \left( x\right) \right] \right) \exp \left( i\int
dy\left[ \overline{\xi }\left( x\right) c^{+}\left( x\right) +\overline{c}%
^{+}\left( x\right) \xi \left( x\right) \right] \right) |\tilde{\Psi}\rangle
\nonumber \\
&&\hspace{-1.5cm}\times \exp \left( \int dx\ dy\ \overline{\xi }(y)\left[
\theta (x_0-y_0)\left\{ c^{-}\left( y\right) ,\overline{c}^{+}\left(
x\right) \right\} -\theta (y_0-x_0)\left\{ c^{+}\left( y\right) ,\overline{c}%
^{-}\left( x\right) \right\} \right] \xi (x)\right) , \\
&=&Z_0^{gh,m}[\xi ,\overline{\xi }]Z_0^{gh,F}[\xi ,\overline{\xi }],
\end{eqnarray}
\noindent where the colour and spinor indices are implicit. As is
clear from the above formulae, the changes in the gluon, quark and
ghost generating functionals with respect to the standard ones in
PQCD are given by the factors:
\begin{eqnarray}
Z_0^{g,m}[j] &=&\langle \tilde{\Psi}|\exp \left( i\int dx\ j^\mu \left(
x\right) A_\mu ^{-}\left( x\right) \right) \exp \left( i\int dy\ j^\mu
\left( x\right) A_\mu ^{+}\left( x\right) \right) |\tilde{\Psi}\rangle ,
\nonumber \\
Z_0^{q,m}[\eta ,\overline{\eta }] &=&\langle \tilde{\Psi}|\exp \left( i\int
dx\left[ \overline{\eta }\left( x\right) \psi ^{-}\left( x\right) +\overline{%
\psi }^{-}\left( x\right) \eta \left( x\right) \right] \right) \exp \left(
i\int dy\left[ \overline{\eta }\left( x\right) \psi ^{+}\left( x\right) +%
\overline{\psi }^{+}\left( x\right) \eta \left( x\right) \right] \right) |%
\tilde{\Psi}\rangle ,  \nonumber \\
Z_0^{gh,m}[\xi ,\overline{\xi }] &=&\langle \tilde{\Psi}|\exp \left( i\int
dx\left[ \overline{\xi }\left( x\right) c^{-}\left( x\right) +\overline{c}%
^{-}\left( x\right) \xi \left( x\right) \right] \right) \exp \left( i\int
dy\left[ \overline{\xi }\left( x\right) c^{+}\left( x\right) +\overline{c}%
^{+}\left( x\right) \xi \left( x\right) \right] \right) |\tilde{\Psi}\rangle
,\   \label{modb}
\end{eqnarray}
which all reduce to unity in the case of the standard PQCD.

Let us now explicitly state the infrared regularization rules of the
expansion to be considered:

\textit{a) The invariant functions }$D$\textit{\ and }$E,$\textit{\ in terms
of which the commutators (anti-commutators) defining the standard Feynman
propagators are expressed, will be assumed to be given by the Nakanishi
infrared regularized expressions }

\begin{eqnarray}
D^{(\pm )}\left( x-y|\sigma ,m\right) &=&\pm \int \frac{dk}{(2\pi )^3}\theta
(\pm k_0-\sigma )\delta (k^2-m^2)\exp (-ik(x-y)),  \nonumber \\
E^{(\pm )}\left( x-y|\sigma \right) \text{ } &=&\frac 12\frac
1{\nabla_{(x-y)}^2}\left( (x_0-y_0)\frac \partial {\partial
(x_0-y_0)}-1\right) \lim_{m\rightarrow 0}D^{(\pm )}\left(
x-y\text{ }|\,\sigma ,m\right) . \label{regnak}
\end{eqnarray}

\textit{b) For the evaluation of the changes in the propagators
produced by the condensation, which are physically related to the
zero-momentum modes of the theory, the unmodified field operators
will be used. In addition, the spatial momenta for which gluon and
quark condensate states are created
will be assumed to lie well within a small neighbourhood of }$\vec{p}=0.$%
\textit{\ The radius }$P$ \textit{of this region will be chosen much smaller
than the Nakanishi parameter }$\sigma $ \textit{in Eq.\ }(\ref{regnak}).

Let us consider in what follows the evaluation of the various Feynman
propagators following the above prescriptions.

\subsection{Regularized Feynman propagators}

\subsubsection{Gluons}

\ The \ Feynman propagator for the gluons is determined by the commutator
between the creation and annihilation parts of their field operators. \
After substituting \ (\ref{sol1}) \ in this commutator, the expression can
be transformed in \ the following way
\begin{eqnarray}
\left[ A_\mu ^{a-}\left( x\right) ,A_\nu ^{b+}\left( y\right) \right]
&=&\delta ^{ab}{\huge (}g_{\mu \nu }D_{+}\left( x-y\,|0\right) -\left(
1-\alpha \right) {\Large (}\partial _\nu ^yD^{(\frac 12)}\left( y\right)
\sum_{\overrightarrow{k}}g_k^{*}(y)\,f_{k,L,\mu }(x)+ \nonumber\\
&&+\partial _\nu ^xD^{(\frac 12)}\left( x\right) \sum_{\overrightarrow{k}%
}g_k(x)f_{k,L,\mu }^{\text{ \ }*}(y){\Large )}{\huge )}, \\
&=&\lim_{\sigma \rightarrow 0}{\Large (}\delta ^{ab}{\large (}g_{\mu \nu
}D_{+}\left( x-y\,|\sigma ,0\right) -\left( 1-\alpha \right) \,\partial _\nu
^y\partial _\mu ^x\,E_{+}\left( x-y\,|\sigma \right) {\large )}{\Large )},
\end{eqnarray}
where it has been defined $D_{\pm }\left( x-y\,|0\right) =\lim_{\sigma
\rightarrow 0}$ $D_{\pm }\left( x-y\,|\sigma ,0\right) .$ In the last line
of this relation, in which the commutator has been expressed in terms of the
invariant functions, the Nakanishi regularization (\ref{regnak}) has been
introduced. \ The functions taken for a finite value of $\sigma ,$ will be
employed in what follows, defining in this way the regularized propagators.

\ \ \ Substituting the above formula in the quadratic form in the gluon
sources defining the Feynman contribution $Z_0^{g,F}$ for the generating
functional in (\ref{fey1}), we now have the form
\begin{align}
S_T& =\int dx\,dy\,\,\theta (y_0-x_0)j^\mu \left( x\right) {\Large [}A_\mu
^{-}\left( x\right) ,\,A_\nu ^{+}\left( y\right) {\Large ]}\,j^\nu (y),
\label{stotal} \\
& =\frac 12\int dx\,dy\,j^{\mu ,a}(x)g_{\mu \nu }\,(\theta
(x_0-y_0)D_{+}\left( x-y\,|\sigma ,0\right) +\theta (y_0-x_0)\text{ }%
D_{+}\left( y-x\,|\sigma ,0\right) \text{\ )}j^{\nu ,a}(y)-  \nonumber \\
-& \frac{\left( 1-\alpha \right) }2\int dx\,dy\,\ \partial _\mu ^xj^{\mu
,a}(x)(\theta (x_0-x_0)E_{+}\left( x-y\,|\sigma \right) +\theta
(y_0-y_0)E_{+}\left( y-x\,|\sigma \right) )\partial _\nu ^yj^{\nu ,a}(y),
\nonumber \\
& =\frac 12\int dx\,dy\,j^{\mu ,a}(x)D_{F\mu \nu }(x-y|\,\sigma )j^{\nu
,a}(y)+  \nonumber \\
& +\frac{\left( 1-\alpha \right) }2\int dx\,dy\,\ \partial _\mu ^xj^{\mu
,a}(x)E_{F\mu \nu }(x-y|\,\sigma )\partial _\nu ^yj^{\nu ,a}(y),  \nonumber
\end{align}
where, in order to shift the derivatives to act on the sources, the
following \ properties have been used\
\begin{align*}
\partial _\nu ^y\partial _\mu ^x(\,E_F(x-y\,|\sigma ))& =\partial _\nu
^y\partial _\mu ^x(\,\theta (x_0-y_0)E_{+}(x-y|\sigma \,)-\theta
(y_0-x_0)E_{-}(x-y\,|\sigma )), \\
& =(\,\theta (x_0-y_0)\partial _\nu ^y\partial _\mu ^xE_{+}(x-y\,|\sigma
)-\theta (y_0-x_0)\partial _\nu ^y\partial _\mu ^xE_{-}(x-y\,|\sigma )),
\end{align*}
in which we had defined $\ \partial _\mu =\nabla _\mu +u_\mu \partial _0$
with $\nabla _\mu \equiv (0,\partial _1,\partial _2,\partial _3)$, $u_\mu
\equiv (1,0,0,0)$, and the final equality follows because the
delta-functions evaluate the factors $(E_{+}(x-y)+E_{-}(x-y))$ at $x_0=y_0,$
where $E_{+}(x-y|\sigma )=-E_{-}(x-y|\sigma )$ is satisfied. Then, the
Feynman propagator contribution to the generating functional of the gluons
takes the form \
\begin{align}
\exp (S_D+S_E)& =\exp \left( \int dx\text{ }dy\,\theta
(y_0-x_0)j(x)[A^{-}\left( x\right) ,\,A^{+}\left( y\right) ]j(y)\right) ,
\nonumber \\
& =\exp \left( \frac 12\int dx\,dy\,j^{\mu ,a}(x)g_{\mu \nu
}D_F(x-y|\,\sigma )j^{\nu ,a}(y)\right) {\small \times }  \nonumber \\
& =\exp \left( {\small +}\frac{\left( 1-\alpha \right) }2\int dx\,dy\,\
\partial _\mu ^xj^{\mu ,a}(x)E_F(x-y\,|\sigma )\partial _\nu ^yj^{\nu
,a}(y)\right) ,
\end{align}
where $D_F(x-y|\,\sigma )$ indicates the Feynman term in the modified gluon
propagator for $\alpha =1$, although it differs in a constant multiplicative
factor from the usual definition.

\ \ \ \ \ \ The generating functional expressed in momentum space for the
gluons can be obtained now by passing the quadratic integrals on the sources
to Fourier representation. The \ following convention for the Fourier
transform \ of any quantity $Q,$ and the Fourier transform of the Heaviside
function will be used \
\begin{align}
Q(x)& =\int \frac{dq}{(2\pi )^4}Q(p)\exp (-iq.x)\equiv \int \frac{dq}{(2\pi
)^4}\mathcal{FT}(Q(x)|x,p)\exp (-iq.x),  \label{fj} \\
\theta (x_0)& =\int_{-\infty }^\infty \frac{dq_0}{2\pi }\frac
i{q_0+i\epsilon }\exp (-iq_0x_0).  \label{fcita}
\end{align}

\ Let us consider first the second term in the argument of the exponential $%
S_E,$ which is the most involved one. After substituting the Fourier
representations (\ref{fj}) for the sources, the Heaviside function (\ref
{fcita}) and\ (\ref{E}), we can obtain
\begin{equation}
S_E=\frac{\left( 1-\alpha \right) }2\int \frac{dk}{(2\pi )^4}k^\mu j_\mu
^a(-k)\left( \mathcal{FT}(\theta (z_0)E_{+}(z|\sigma )|z,k)+\mathcal{FT}%
(\theta (z_0)E_{+}(z|\sigma )|z,-k)\right) k^\nu j_\nu ^a(k).
\label{se}
\end{equation}

The first Fourier transform in the integrand of (\ref{se}), after
substituting the definition (\ref{regnak}) of the $E$ function and
the Fourier representation (\ref{fcita}) of $\theta (z_0),$ can be
obtained in the form

\[
\mathcal{FT}(\theta (z_0)E_{+}(z|\sigma )|z,k)=\frac{i\,\theta (|%
\overrightarrow{k}|-\sigma )\text{ }(2|\overrightarrow{k}|-k_0)}{-4|%
\overrightarrow{k}|^3(k_0-|\overrightarrow{k}|+i\epsilon )^2}.
\]

Since \ the second Fourier transform in the integrand of
(\ref{se}) \ is simply the first one, \ but with the \ integration
momentum $k$ argument replaced by $-k$, adding the two terms leads
to
\begin{align*}
S_E& =\frac{\left( 1-\alpha \right) }2\int \frac{dk}{(2\pi )^4}k^\mu j_\mu
^a(-k){\Huge (}\frac{i\,\theta (|\overrightarrow{k}|-\sigma )\text{ }(2|%
\overrightarrow{k}|-k_0)}{-4|\overrightarrow{k}|^3(k_0-|\overrightarrow{k}%
|+i\epsilon )^2}+ \\
& +\frac{i\,\theta (|\overrightarrow{k}|-\sigma )\text{ }(2|\overrightarrow{k%
}|+k_0)}{-4|\overrightarrow{k}|^3(-k_0-|\overrightarrow{k}|+i\epsilon )^2}%
{\Huge )}k^\nu j_\nu ^a(k). \\
& =-\frac{\left( 1-\alpha \right) }2\int \frac{dk}{(2\pi )^4}k^\mu j_\mu
^a(-k)\frac{i\,\theta (|\overrightarrow{k}|-\sigma )\text{ }}{(k^2+i\epsilon
)^2}k^\nu j_\nu ^a(k).
\end{align*}
\

\ For the \ terms in (\ref{se}) associated to the invariant function \ $%
D_{+},$ \ a similar but much simpler calculation \ thanks to the absence of
dipole fields, allows us to obtain
\[
S_D=\int \frac{dk}{(2\pi )^4}j_\mu ^a(-k)\frac{i\,\theta (|\overrightarrow{k}%
|-\sigma )\text{ }}{2(k^2+i\epsilon )}g^{\mu \nu }j_\nu ^a(k),
\]
which inserted  in \ (\ref{fey1}) gives, for the \ regularized gluon \
Feynman \ factor in the free generating functional, the \ expression \
\[
Z_0^{g,F}[j]=\exp \left( \frac i2\int \frac{dk}{(2\pi )^4}\,j^{\mu ,a}(-k)%
\frac{\theta (|\overrightarrow{k}|-\sigma )}{(k^2+i\epsilon )}\left( g_{\mu
\nu }-\frac{\left( 1-\alpha \right) k_\mu k_\nu }{(k^2+i\epsilon )}\right)
j^{\nu ,a}(k)\right) .
\]

\subsubsection{Ghosts and \ Quarks}

\ In a similar way, the regularized ghost and quark propagators (the quark
taken in the massless limit $m\rightarrow 0)$ can be obtained in the forms \
\begin{align}
Z_0^{q,F}[\eta ,\overline{\eta }]& =\exp \left( i\int \frac{dk}{(2\pi )^4}%
\overline{\eta }(-k)\frac{-\theta (|\overrightarrow{k}|-\sigma )\,k_\mu
\gamma ^\mu }{k^2+i\epsilon \ }\eta (k)\right) ,  \label{quarfeyprop} \\
Z_0^{gh,F}[\xi ,\overline{\xi }]& =\exp \left( i\int \frac{dk}{(2\pi )^4}%
\overline{\xi }(-k)\frac{-i\theta (|\overrightarrow{k}|-\sigma )\,\delta
^{ab}}{k^2+i\epsilon }\xi (k)\right) .  \label{ghostfeyprop}
\end{align}

\ Therefore, the Nakanishi regularization leads to the vanishing of all the
Feynman propagator terms within a tube containing $k=0$ and defined by the
momenta having modulus of their spatial component \ smaller that $\sigma ,$\
for arbitrary values of the \ temporal component $k_0$. In this way, as will
be seen further in the discussion, the adoption of this \ regularization
procedure naturally leads to the cancellation of a great number of the
singularities in the perturbative expansion.

\subsection{ Condensation-induced changes in the propagators}

The effects on the propagators of the gluons and quarks  created
by the condensation of gluons, quarks and ghost in the free ground
state, will be considered in this subsection. Let us separately
examine the two cases.

\subsubsection{Gluons}

Taking into account  expressions (\ref{sol1}), the gluon
annihilation and creation parts in (\ref{modb}) can be explicitly
written as
\begin{eqnarray}
A_\mu ^{a+}\left( x\right) &=&\sum\limits_{\vec{k}}\left(
\sum\limits_{\lambda =1,2}A_{\vec{k},\lambda }^af_{k,\mu }^\lambda \left(
x\right) +A_{\vec{k}}^{L,a}f_{k,L,\mu }\left( x\right) +B_{\vec{k}}^a\left[
f_{k,S,\mu }\left( x\right) +(1-\alpha )\left( A\left(k_0,x_0\right)
f_{k,L,\mu }\left( x\right) +B_\mu \left( k,x\right) \right) \right] \right)
,  \label{4.11} \\
A_\mu ^{a-}\left( x\right) &=&\sum\limits_{\vec{k}}\left(
\sum\limits_{\lambda =1,2}A_{\vec{k},\lambda }^{a+}f_{k,\mu }^{\lambda
*}\left( x\right) +A_{\vec{k}}^{L,a+}f_{k,L,\mu }^{*}\left( x\right) +B_{%
\vec{k}}^{a+}\left[ f_{k,S,\mu }^{*}\left( x\right) +(1-\alpha
)\left( A^{*}\left(k_0,x_0\right) f_{k,L,\mu }^{*}\left( x\right)
+B_\mu ^{*}\left( k,x\right) \right) \right] \right),  \nonumber
\end{eqnarray}
where $A\left( k_{0},x_{0}\right)$ and $B_{\mu}\left(k,x\right)$ are defined
as
\[
A\left( k_{0},x_{0}\right) \equiv\frac{1}{2\left\vert \vec{k}\right\vert ^{2}%
}\left( ik_{0}x_{0}+1/2\right) ,\qquad B_{\mu}\left( k,x\right) \equiv\frac{%
ik_{0}\delta_{\mu0}}{2\left\vert \vec{k}\right\vert ^{2}}g_{k}(x).
\]

On arriving to (\ref{4.11}), the term of
$\partial_{\mu}D_{(x)}^{(1/2)}
g_{k}(x)$ in (\ref{sol1}) was evaluated by using the wave packets (\ref{paq}%
) and polarization vectors (\ref{polar}) with the result
\begin{equation}
\partial_{\mu}D_{(x)}^{(1/2)}g_{k}(x)=\partial_{\mu}1/2(\nabla^{2}
)^{-1}(x_{0}\partial_{0}-1/2)\frac{1}{\sqrt{2Vk_{0}}} \exp\left(-ikx%
\right)=A\left( k_{0},x_{0}\right) f_{k,L,\mu}\left( x\right)+B_{\mu}\left(
k,x\right),  \label{1}
\end{equation}
which defined $A\left( k_{0},x_{0}\right)$ and
$B_{\mu}\left(k,x\right)$. In a similar way, for
$\partial_{\mu}D_{(x)}^{(1/2)}g_{k}^{\ast}(x)$ we obtained
\begin{equation}
\partial_{\mu}D_{(x)}^{(1/2)}g_{k}^{\ast}(x)=A^{\ast}\left( k_{0}
,x_{0}\right) f_{k,L,\mu}^{\ast}\left( x\right) +B_{\mu}^{\ast}\left(
k,x\right) .  \label{2}
\end{equation}

As commented in Appendix A, the transverse term is not modified by
the generalization of the theory for arbitrary values of $\alpha$.
Then, its contribution is the same as calculated previously
\cite{PRD,tesis}. It is important to recall that thanks to the
diagonal block structure commutation relations (\ref{commu}) the
calculations are done by decomposing the expression (\ref{modb})
in products of contributions coming from separate modes associated
to the diagonal blocks.

The result obtained in Refs.\ \cite{PRD,tesis} for the
transverse-mode contributions, for $\left| C_\lambda \left(
p_i\right) \right| <1$, is
\begin{eqnarray}
&&\frac 1N\langle 0\mid \exp \left( \sum_{\lambda =1,2}\frac{C_\lambda
^{*}\left( P\right) }2A_{\vec{p}_i,\lambda }^aA_{\vec{p}_i,\lambda
}^a\right) \exp \left\{ i\int dxJ^{\mu ,a}\left( x\right)
\sum\limits_{\lambda =1,2}A_{\vec{p}_i,\lambda }^{a+}f_{p_i,\mu }^{\lambda
*}\left( x\right) \right\}   \nonumber \\
&&\times \exp \left\{ i\int dxJ^{\mu ,a}\left( x\right) \sum\limits_{\lambda
=1,2}A_{\vec{p}_i,\lambda }^af_{p_i,\mu }^\lambda \left( x\right) \right\}
\exp \left( \sum_{\lambda =1,2}\frac{C_\lambda \left( P\right) }2A_{\vec{p}%
_i,\lambda }^{a+}A_{\vec{p}_i,\lambda }^{a+}\right) \mid 0\rangle ,
\nonumber \\
&=&\exp \left\{ -\sum\limits_{\lambda =1,2}\left( J_{p_i,\lambda }^a\right)
^2\frac{\left( C_\lambda \left( P\right) +C_\lambda ^{*}\left( P\right)
+2\left| C_\lambda \left( P\right) \right| ^2\right) }{2\left( 1-\left|
C_\lambda \left( P\right) \right| ^2\right) }\right\} ,  \label{Transv}
\end{eqnarray}
where the normalization factor has been cancelled. It should be
recalled that the momenta $\vec{p}_i$ are very small and the
sources are assumed to be located in a finite spatial region. The
following simplified notation was also introduced in
\cite{PRD,tesis}:
\[
j_{p_i,\lambda }^a=\int \frac{dx}{\sqrt{2Vp_{i0}}}j^{\mu ,a}\left( x\right)
\epsilon _{\lambda ,\mu }\left( p_i\right) .
\]

The calculation of longitudinal and scalar-mode contributions is,
on this occasion, more elaborate, and a brief exposition of it can
be found in Appendix A; the result obtained for $\left\vert
C_{3}\left( \alpha,P\right) \right\vert <1$ is
\begin{eqnarray}
&&\exp \left\{ -\int \frac{dxdy}{2Vp_{i0}}J^{\mu ,a}\left( x\right) J^{\nu
,a}\left( y\right) \left[ \left( \frac{C_3\left( \alpha ,P\right)
+C_3^{*}\left( \alpha ,P\right) +2\left| C_3\left( \alpha ,P\right) \right|
^2}{\left( 1-\left| C_3\left( \alpha ,P\right) \right| ^2\right) }%
\right)\times \right. \right.  \nonumber \\
&&\qquad \qquad\qquad \times \left( \epsilon _{S,\mu }\left( p_i\right)
\epsilon _{L,\nu }\left( p_i\right) +\frac{\left( 1-\alpha \right) }{4\left|
\vec{p}_i\right| ^2} \left[ \epsilon _{L,\mu }\left( p_i\right)
+i2p_{i0}\delta _{\mu 0}\right] \epsilon _{L,\nu }\left( p_i\right) \right) +
\nonumber \\
&&\qquad\qquad \left. \left. +\left( \frac{D^{*}\left( \alpha ,P\right)
\left( C_3\left( \alpha ,P\right) +1\right) ^2+D\left( \alpha ,P\right)
\left( C_3^{*}\left( \alpha ,P\right) +1\right) ^2}{\left( 1-\left|
C_3\left( \alpha ,P\right) \right| ^2\right) ^2}\right) \epsilon _{L,\mu
}\left( p_i\right) \epsilon _{L,\nu }\left( p_i\right) \right] \right\}
\label{LonSca} \\
&&\times \langle 0 \mid \exp \left( C_3^{*}\left( \alpha ,P\right) B_{\vec{p}%
_i}^aA_{\vec{p}_i}^{L,a}+D^{*}\left( \alpha ,P\right) B_{\vec{p} _i}^aB_{%
\vec{p}_i}^a\right) \exp \left( C_3\left( \alpha ,P\right) B_{\vec{p}
_i}^{a+}A_{\vec{p}_i}^{L,a+}+D\left( \alpha ,P\right) B_{\vec{p}_i}^{a+}B_{%
\vec{p}_i}^{a+}\right) \mid 0\rangle,  \nonumber
\end{eqnarray}
where the normalization factor (last line) is equal to one. Now
inserting expressions (\ref{Transv}) and (\ref{LonSca}) in
 (\ref{modb}), performing some algebraic manipulations [keeping in
mind the properties of the polarization vectors defined in
(\ref{polar})], assuming $C_{1}\left( P\right)
=C_{2}\left(P\right)$, and introducing the contribution of all
momenta $\vec{p}_i$ ($|\vec{p}_i|=P$), the following expression is
obtained:
\begin{eqnarray}
&&\exp \left\{ \int \frac{dxdy}{2V}J^{\mu,a}\left( x\right) J^{\nu,a}\left(
y\right) \sum\limits_{\vec{p}_i, \left| \vec{p}_i\right| =P} \frac
1{p_{i0}}\left[ \left( \frac{C_1\left( P\right) +C_1^{*}\left( P\right)
+2\left| C_1\left( P\right) \right| ^2}{2\left( 1-\left| C_1\left( P\right)
\right| ^2\right) }\right) g_{\mu \nu }+\right. \right.  \nonumber \\
&&\text{ \ }\qquad +\left( \frac{C_3\left( \alpha,P\right) +C_3^{*}\left(
\alpha,P\right) +2\left| C_3\left( \alpha,P\right) \right| ^2}{\left(
1-\left| C_3\left( \alpha,P\right) \right| ^2\right) }-\frac{C_1\left(
P\right) +C_1^{*}\left( P\right) +2\left| C_1\left( P\right) \right| ^2}{%
\left( 1-\left| C_1\left( P\right) \right| ^2\right) }\right) \frac{\bar{p}
_{i\mu }p_{i\nu }}{2\left| \vec{p}_i\right| ^2}  \nonumber \\
&&\text{ \qquad }+\left( \frac{C_3\left( \alpha,P\right) +C_3^{*}\left(
\alpha ,P\right) +2\left| C_3\left( \alpha,P\right) \right| ^2}{\left(
1-\left| C_3\left( \alpha,P\right) \right| ^2\right) }\right) \frac{\left(
1-\alpha \right) }{4\left| \vec{p}_i\right| ^2}\left( p_{i\mu }p_{i\nu
}-2p_{i0}\delta _{\mu 0}p_{i\nu }\right)  \nonumber \\
&&\text{ \qquad }+\left. \left. \frac{D^{*}\left( \alpha,P\right) \left(
C_3\left( \alpha,P\right) +1\right) ^2+D\left( \alpha,P\right) \left(
C_3^{*}\left( \alpha,P\right) +1\right) ^2}{\left( 1-\left| C_3\left(
\alpha,P\right) \right| ^2\right) ^2}p_{i\mu }p_{i\nu }\right) \right\}.
\label{mod1}
\end{eqnarray}
It is important to notice at this point that the combinations of
$C_1\left( P\right)$, $C_3\left( \alpha ,P\right)$, and
$D^{*}\left( \alpha,P\right)$ in Eq.\ (\ref{mod1}) are real. So
that even if these parameters are complex, only their real parts
contribute to the propagator. After this observation, they will be
considered real in what follows.

In expression (\ref{mod1}) we take the thermodynamic limit ($V \rightarrow
\infty$), replace sums by integrals,
\begin{equation}
\frac 1V \sum\limits_{\vec{p}_i, \left| \vec{p}_i\right| =P}=\frac 1{\left(
2\pi \right) ^3}\int\limits_0^\infty dp\ p^2\delta\left( p -
P\right)\int\limits_0^\pi \sin\theta d\theta \int\limits_0^{2\pi }d\varphi,
\label{integr}
\end{equation}
and perform the integrations.

As a second step, we proceed to take the limit $P\rightarrow 0$, for that we
consider that the parameters introduced in the modified vacuum state to be
\begin{eqnarray}
C_1\left( P\right) &\sim &1-\frac{C_1}2P,\quad C_1>0,  \nonumber \\
C_3\left( \alpha,P\right) &\sim &1-\frac{C_3\left( \alpha \right) }2P,\quad
C_3\left( \alpha \right) >0,  \nonumber \\
D\left( \alpha,P\right) &\sim &D\left( \alpha \right).  \label{cond}
\end{eqnarray}

After that, the result obtained for Eq.\ (\ref{mod1}) in the limit $%
P\rightarrow 0$ is
\begin{eqnarray}
&&\exp \left\{ \int dxdyJ^{\mu,a}\left( x\right) J^{\nu ,a}\left( y\right)
\left[ \frac{2g_{\mu \nu }}{\left( 2\pi \right) ^2C_1}+\frac 2{3\left( 2\pi
\right) ^2}\left( \frac 1{C_3\left( \alpha \right) }-\frac 1{C_1}\right)
\left( g_{\mu \nu }+2\delta _{\mu 0}\delta _{\nu 0}\right) \right. \right.
\nonumber \\
&&\text{ \qquad \qquad \qquad }\left. \left. -\frac{\left( 1-\alpha \right)
}{\left( 2\pi \right) ^23C_3\left( \alpha \right) }\left( g_{\mu \nu
}+2\delta _{\mu 0}\delta _{\nu 0}\right) -\frac{4D\left( \alpha \right) }{%
\left( 2\pi \right) ^23\left[ C_3\left( \alpha \right) \right] ^2}\left(
g_{\mu \nu }-4\delta _{\mu 0}\delta _{\nu 0}\right) \right) \right\}.
\label{mod2}
\end{eqnarray}

In order to make explicit the Lorentz invariance, we fix $D\left(
\alpha \right)$, such that the terms proportional to $\delta _{\mu
0}\delta _{\nu 0}$ cancel out. Then $D\left( \alpha \right)$ is
determined by the expression
\[
D\left( \alpha \right) =\frac{\left[ C_3\left( \alpha \right) \right] ^2}{4}%
\left[ \frac 1{C_1}-\frac{\left( 1+\alpha \right) }{2C_3\left( \alpha
\right) } \right],
\]
and (\ref{mod2}) takes the form
\begin{equation}
\exp \left\{ \int dxdyJ^{\mu,a}\left( x\right) J^{\nu ,a}\left( y\right)
\frac{g_{\mu \nu }}{\left( 2\pi \right) ^2}\left( \frac 1{C_1}+ \frac{\left(
1+\alpha \right) }{2C_3\left( \alpha \right) }\right) \right\}.  \label{mod3}
\end{equation}

In expression (\ref{mod3}) we notice that the term in the
exponential is real and non-negative. This is because $C_1>0$,
$C_3\left( \alpha \right) >0$ and $\alpha \geq 0$ ($\alpha $ most
be non-negative for the convergence of the Gaussian integral in
which it was introduced in the path integral formalism. It can be
stressed at this point, that, if could be possible  justifying the
validity of the recursive solution as an analytical extension in
the parameters, it looks possible to trace a connection with the
approach of Refs. \cite{Munczek,Burden,pavel}). As a final step, we chose the parameter $%
C_3\left( \alpha \right) $ in such a way that the determined
modification does not depend on the gauge parameter $\alpha $, as
it was selected in an earlier work \cite{1995}. Then $C_3\left(
\alpha \right) =\frac 12KC_1\left( 1+\alpha \right), $ with $K$ a
positive constant.

Considering the above remarks and defining in the generating
functional modification an alternative non-negative constant
$C_g=\frac{2\left( 2\pi \right)^2}{C_1}\left(1+ \frac{1}K\right)$
(which we call condensate parameter) for further convenience,
expression (\ref{mod3}) takes the form
\[
\exp \left\{ \int dxdy\sum\limits_{a=1}^8J^{\mu,a}\left( x\right)
J^{\nu,a}\left( y\right) \frac{g_{\mu \nu }C_g}{\left( 2\pi \right) ^42}
\right\}.
\]

From the above construction, the parameter $D\left( \alpha \right)
=D\left(1+\alpha \right)^2$, where $D$ is an arbitrary real
constant (equal to zero for $K=1$), and the vacuum in the
zero-momentum limit has the form
\begin{equation}
\mid \Psi \rangle =\exp \left[ \sum\limits_{a=1}^8\frac
12A_{0,1}^{a+}A_{0,1}^{a+}+\frac
12A_{0,2}^{a+}A_{0,2}^{a+}+B_0^{a+}A_0^{L,a+}+i\overline{c}
_0^{a+}c_0^{a+}+D\left(1+\alpha \right)^2B_0^{a+}B_0^{a+}\right] \mid
0\rangle,  \label{Vacuum3}
\end{equation}
from which the state selected in the previous works
\cite{PRD,tesis}, corresponds
to the particular case $\alpha=1$, $C_3\left( \alpha=1 \right) =C_1$ and $%
D\left(\alpha=1 \right) =0$.

Thus, it follows that the vacuum state introduced in
(\ref{Vacuum3}) modifies the usual perturbation theory only
through a change in the gluon propagator which now takes the form
\begin{equation}
D_{\mu \nu }^{ab}(x-y)=\int \frac{dk}{\left( 2\pi \right) ^4} \delta
^{ab}\left[ \frac{\theta(|\vec{k}|-\sigma)} {k^{2}+i\epsilon} \left( g_{\mu
\nu }-\left( 1-\alpha \right) \frac{k_\mu k_\nu }{k^2+i\epsilon}\right)
-iC_g\delta \left( k\right) g_{\mu \nu }\right] \exp \left\{ -ik\left(
x-y\right) \right\}.  \label{propag}
\end{equation}
in which the first term is the usual Feynman propagator, but now
including an infrared regularization.

\subsubsection{Ghosts}

The change in the ghost generating functional is the same as
calculated in Refs. \cite{PRD,tesis}. This is because the ghost
sector in the vacuum
state remains unchanged. As  was mentioned in these references for the value $%
C_{3}\left( \alpha,0\right) =1$ as it was fixed here, there is no
modification of the ghost sector and its generating functional and
propagator remain the same as for the usual PQCD.

\subsubsection{Quarks}

Let us now consider  the modification in the quark propagator
produced by the presence of its condensate. The factor
corresponding to quarks in the generating functional (\ref{zt})
can be written as
\begin{eqnarray}
Z_0^{q,m}[\eta ,\overline{\eta }] &=&\langle 0\,|\exp \left(
\sum_{s=1,2}\sum\limits_{p_i,q_i,\left| p_i,q_i\right| =P<\sigma }C_{\vec{p}%
_i,\vec{q}_i}^{*}\ a_{\vec{p}_i}^sb_{\vec{q}_i}^s\right) \exp \left\{ i\int
dx\left[ \overline{\eta }\left( x\right) \psi ^{-}\left( x\right) +\overline{%
\psi }^{-}\left( x\right) \eta \left( x\right) \right] \right\}   \nonumber
\\
&&\times \exp \left\{ i\int dy\left[ \overline{\eta }\left( x\right) \psi
^{+}\left( x\right) +\overline{\psi }^{+}\left( x\right) \eta \left(
x\right) \right] \right\} \exp \left(
\sum_{s=1,2}\sum\limits_{p_i,q_i,\left| p_i,q_i\right| =P<\sigma }C_{\vec{p}%
_i,\vec{q}_i}\ b_{\vec{q}_i}^{s+}a_{\vec{p}_i}^{s+}\right) |0\rangle ,
\label{zv} \\
&=&\langle 0\,|\exp \left[ C_{ff^{\prime }}^{*}a_f^sb_{f^{\prime }}^s\right]
\exp \left\{ -\sigma _fa_f^{+}+\overline{\sigma }_fb_f^{+}\right\} \exp
\left\{ -\sigma _f^{*}a_f+\overline{\sigma }_f^{*}b_f\right\} \exp \left[
C_{ff^{\prime }}b_{f^{\prime }}^{s+}a_f^{s+}\right] |\,0\rangle ,  \nonumber
\\
&=&Z_0^{q,m}[\sigma ,\overline{\sigma }]\equiv Z_0^{q,m}[\text{v}]\text{ \ ,
v}\equiv (\sigma ,\overline{\sigma }),\,\ \ |\,q\rangle =\exp \left[
C_{ff^{\prime }}b_{f^{\prime }}^{s+}a_f^{s+}\right] |\,0\rangle ,  \nonumber
\end{eqnarray}
where the quark-field operators were expressed as sums over the
creation and annihilation components. In order to simplify the
discussion the following compact notation has been employed
$f,f^{\prime }=\vec{p}_i,i=1,2,3,\ldots ,N.$ The new indices, $f$
and $f^{\prime },$ correspond to the values of the momenta for
which condensate states are created. The spinor index $s=1$ or $2$
have been omitted, since the discussion can be done almost up to
the end  for each value of $s$, thanks to the commutativity of the
creation operators for different indices. The coefficients of the
quark creation and annihilation operators $a^{\pm }$ and $b^{\pm
}$ are expressed as
\begin{equation}
\sigma _f\equiv \sigma _{\vec{k}}^s=\frac i{\sqrt{V}}\int dx\,\overline{u}_{%
\vec{k}}^s\,\eta (x)\exp (ikx),\qquad \overline{\sigma }_f\equiv \overline{%
\sigma }_{\vec{k}}^s=\frac i{\sqrt{V}}\int dx\,\overline{\eta }(x)\,%
\overline{v}_{\vec{k}}^s\,\exp (ikx),  \label{sigmas}
\end{equation}
where $u$ and $v$ are the usual spinor solutions of the Dirac equation
defined in Appendix A.

Let us now define the transformations $U_1^{-1}=\exp (-\sigma _fa_f^{+}+%
\overline{\sigma }_fb_f^{+})$ and $U_2=\exp (-\sigma _f^{*}a_f+\overline{%
\sigma }_f^{*}b_f),$ which acting upon the creation and
annihilation operators through a similarity, lead to
\begin{eqnarray*}
U_1b_fU_1^{-1} &=&b_f-\overline{\sigma }_f,\qquad U_1a_fU_1^{-1}=a_f+\sigma
_f, \\
U_2b_f^{+}U_2^{-1} &=&b_f^{+}+\overline{\sigma }_f^{*},\qquad
U_1a_f^{+}U_1^{-1}=a_f^{+}-\sigma _f^{*}.
\end{eqnarray*}

After also consider $\langle 0\,|U_1=\langle 0|$ and $U_2|0\rangle =|0\rangle $%
; it follows that
\begin{eqnarray}
Z_0^{q,m}[\text{v}] &=&\langle 0\,|\exp \left[ C_{f\,f^{\prime
}}^{*}(a_f^{}+\sigma _f)(b_{f^{\prime }}^{}-\overline{\sigma }_{f^{\prime
}}^{})\right] \exp \left[ C_{f\,f^{\prime }}(b_{f^{\prime }}^{s+}+\overline{%
\sigma }_{f^{\prime }}^{*})(a_f^{s+}-\sigma _f^{*})\right] |\,0\rangle ,
\label{zv1} \\
&=&\langle q|\exp \left[ -C_{f\,f^{\prime }}^{*}\sigma _f\overline{\sigma }%
_{f^{\prime }}^{}-C_{f\,f^{\prime }}\overline{\sigma }_{f^{\prime
}}^{*}\sigma _f^{*}\right] \exp \left[ -a_f\,C_{f\,f^{\prime }}\overline{%
\sigma }_{f^{\prime }}+\sigma _fC_{_{f\,f^{\prime }}}b_{f^{\prime }}\right]
\exp \left[ -a_f^{*}\,C_{f\,f^{\prime }}\overline{\sigma }_{f^{\prime
}}^{*}+\sigma _f^{*}C_{_{f\,f^{\prime }}}b_{f^{\prime }}^{+}\right]
|\,q\rangle .  \nonumber
\end{eqnarray}

It is now possible to employ the following identity for operators
that commute with their commutators
\begin{equation}
\exp (A)\exp (B)=\exp (B)\exp (A)\exp ([A,B]),  \label{haus}
\end{equation}
to change the order of the exponential operators appearing in the
mean value on the quark `squeezed´ state in Eq.\ (\ref{zv1}). The
commutator between the two arguments of those exponentials can be
calculated to be
\[
\left[ -a_f\,C_{f\,f^{\prime }}\overline{\sigma }_{f^{\prime }}+\sigma
_fC_{_{f\,f^{\prime }}}b_{f^{\prime }},-a_f^{*}\,C_{f\,f^{\prime }}\overline{%
\sigma }_{f^{\prime }}^{*}+\sigma _f^{*}C_{_{f\,f^{\prime }}}b_{f^{\prime
}}^{+}\right] =-\left[ \overline{\sigma }_{f^{\prime }}C_{f^{\prime
}\,f}^{T*}C_{f\,f^{\prime \prime }}\overline{\sigma }_{f^{\prime \prime
}}+\sigma _fC_{f\,f^{\prime }}^{*}\,C_{f^{\prime }\,f^{\prime \prime
}}^T\sigma _{f^{\prime \prime }}^{*}\right] ,
\]
which, after being considered together with (\ref{haus}), allows
us  to write  the quark modification as
\begin{eqnarray}
Z_0^{q,m}[\text{v}] &=&\exp \left[ -\sigma \widehat{C}\overline{\sigma }-%
\overline{\sigma }^{*}\widehat{C}\sigma ^{*}-\overline{\sigma }\widehat{C}^T%
\widehat{C}\overline{\sigma }^{*}-\sigma \widehat{C}^T\widehat{C\,}\sigma
^{*}\right] Z\left[ -\widehat{C\,}\overline{\sigma }^{*},\widehat{C}^T\sigma
^{*}\right] ,  \label{iter} \\
&=&\exp \left[ \frac 12\text{v}_1R\,\text{v}_1-\frac 12\text{v}_1^{*}R\,%
\text{v}_1^{*}+\text{v}_1R^2\,\text{v}_1^{*}\right] Z_q\left[ R\,\mathcal{K}%
\text{v}_1\right] ,  \nonumber
\end{eqnarray}
where it was assumed that $C_{ff^{\prime }}$ is real, and the following
matrix notation was introduced
\begin{eqnarray*}
\sigma \widehat{C}\overline{\sigma } &\equiv &a_f\,C_{f\,f^{\prime }}%
\overline{\sigma }_{f^{\prime }},\quad (\widehat{C}^T)_{ff^{\prime }}=(%
\widehat{C})_{f^{\prime }f},\quad \text{v}_1=%
\begin{pmatrix}
\sigma\\
\overline{\sigma}
\end{pmatrix}
,\quad R=%
\begin{pmatrix}
0 & -\widehat{C}\\
\widehat{C} & 0
\end{pmatrix}
,\quad  \\
\mathcal{K}\text{v}_1 &=&%
\begin{pmatrix}
\sigma^{\ast}\\
\overline{\sigma}^{\ast}
\end{pmatrix}
,\quad \mathcal{K}^2=I,
\end{eqnarray*}
and $\mathcal{K}$ is simply the complex conjugate of a vector
expressed in matrix notation. Relation (\ref{iter}), after being
applied $n$ times, leads to the following expression:
\[
Z_0^{q,m}[\text{v}]=\exp \left\{ \frac 12\text{v}_1R\left(
\sum_{m=0}^n\,R^{2m}\right) \text{v}_1-\frac 12\text{v}_1^{*}R\left(
\sum_{m=0}^n\,R^{2m}\right) \text{v}_1^{*}+\text{v}_1R^2\,\left(
\sum_{m=0}^n\,R^{2m}\right) \text{v}_1^{*}\right\} \,Z_q\left[ \left(
\mathcal{K}R\right) ^{n+1}\text{v}_1\right] .
\]

However, the sums of powers of the matrix $R$ can be written in
the form
\[
\sum_{m=0}^{n}\,R^{2m}=
\begin{pmatrix}
\sum_{m=0}^{n}(-1)^{m}\,(\widehat{C}\widehat{C}^{T})^{m} & 0\\
0 & \sum_{m=0}^{n}\,(-1)^{m}(\widehat{C}^{T}\widehat{C})^{m}
\end{pmatrix},
\]
which gives in the limit $n\rightarrow\infty$, assuming that all
the eigenvalues of the symmetric matrix $R^{2}$ are smaller than
1:
\[
\sum_{m=0}^{\infty}\,R^{2m}=
\begin{pmatrix}
\frac{1}{1+\widehat{C}\widehat{C}^{T}} & 0\\
0 & \frac{1}{1\widehat{+C}^{T}\widehat{C}}
\end{pmatrix}
=\frac{1}{1-R^{2}}.
\]

Hence, the modification of the quark propagator takes the form \
\begin{equation}
Z_0^{q,m}[\text{v}]=\exp \left\{ \frac 12\text{v}_1\frac R{1-R^2}\text{v}%
_1+\frac 12\text{v}_1^{*}\frac R{1-R^2}\text{v}_1^{*}+\text{v}_1\frac{R^2}{%
1-R^2}\text{v}_1^{*}\right\} \,Z_q[0],  \label{expoa}
\end{equation}
where the r.h.s. of the generating functional is zero because $%
\lim_{n\rightarrow 0}(R^{n+1})=0$, if all the eigenvalues of $R^2$
are assumed to be smaller than the unity. For the argument of the
exponential above it follows
\[
\text{v}_1\frac{R^2}{1-R^2}\text{v}_1^{*}=-\sigma \frac{\widehat{C}\widehat{C%
}^T}{1+\widehat{C}\widehat{C}^T}\sigma ^{*}-\overline{\sigma }\frac{\widehat{%
C}^T\widehat{C}}{1+\widehat{C}^T\widehat{C}}\overline{\sigma }^{*}.
\]
Now, assuming that the matrix $C$ is antisymmetric and also that
it  satisfies $\ \widehat{C\text{ }}^T\widehat{C}=K\widehat{I}$
(with $K$ including also possible negative values), it follows
that the sum of the first two terms in the argument of the
exponential in (\ref{expoa}) vanishes. The one remaining take the
form
\[
v_1\frac{R^2}{1-R^2}v_1^{*}=\frac K{1-K}\left( \sigma \sigma ^{*}+\overline{%
\sigma }\overline{\sigma }^{*}\right) =\frac K{1-K}\left( \sigma _f\sigma
_f^{*}+\overline{\sigma _f}\overline{\sigma }_f^{*}\right) =\frac
K{1-K}\sum_{\vec{k}}\left( \sigma _{\vec{k}}^s\sigma _{\vec{k}}^{s*}+\sigma
_{\vec{k}}^s\overline{\sigma }_{\vec{k}}^{s*}\right)
\]
where the index $s=1,2$ takes the fixed value that has been used
for the whole evaluation. Now employing the definitions (\ref{sigmas}) of $\sigma $ and $%
\overline{\sigma }$ and the identities
\[
\sum_{s=1,2}v_{\vec{k}}^s\overline{v}_{\vec{k}}^s=\left( \gamma ^\mu p_\mu
-m\right) ,\qquad \sum_{s=1,2}u_{\vec{k}}^s\overline{u}_{\vec{k}}^s=\left(
\gamma ^\mu p_\mu +m\right) ,
\]
and assuming that the momenta arguments $p_i$ are taken at their
vanishing value in the limit $L\rightarrow \infty $, it follows
that
\begin{eqnarray*}
\frac K{1-K}\sum_{\vec{k},s=1,2}\left( \sigma _{\vec{k}}^s\sigma _{\vec{k}%
}^{s*}+\sigma _{\vec{k}}^s\overline{\sigma }_{\vec{k}}^{s*}\right)
&=&-\frac K{1-K}\frac mV\sum_{\vec{k}}\int dx\,dy\,\overline{\eta }(x)\,\eta
(y)\,\exp (ik(x-y)), \\
&=&-\frac K{1-K}\frac mVN\int dx\,dy\,\overline{\eta }(x)\,\eta (y)\,\exp
(ik(x-y)),
\end{eqnarray*}
where the number $N$ of the momenta for the created particles and
antiparticles appears in the last line from  the assumption that
for small momenta the integral appearing is almost independent of
$k$. Fixing $K$ as given by
\[
K=\frac 1{1-\frac{m(L)N(2\pi )^4}{C_qL^3}},
\]
and considering (\ref{quarfeyprop}) allows us to write for the
change in the quark generating functional and propagators
\begin{eqnarray}
Z_0^{q,m}[\eta ,\overline{\eta }] &=&\exp \left( C_q\int dx\,dy\,\overline{%
\eta }(x)\eta (y)\right) Z_q[0], \\
Z_0^{q,F}[\eta ,\overline{\eta }] &=&\exp \left( \int dx\,dy\,\,\overline{%
\eta }(x)G_F(x-y)\eta (y)\right) Z_q[0], \\
Z_0^q[\eta ,\overline{\eta }] &=&\exp \left( \int dx\,dy\,\,\overline{\eta }%
(x)G^q(x-y)\eta (y)\right) Z_q[0], \\
G^q(x-y) &=&\int \frac{dk}{\left( 2\pi \right) ^4}\left[ -\frac{i\theta (|%
\vec{k}|-\sigma )\gamma ^\nu k_\nu }{k^2+i\epsilon }+C_qI\,\delta \left(
k\right) \right] \exp (-ik\left( x-y\right) ).
\end{eqnarray}

\section{Regularization of the singular terms}

A main technical difficulty for the implementation of the
expansion proposed in Refs.\ \cite{1995,PRD,epjc,Hoyer,jhep} is
that the modifications of the usual free propagator terms (to be
called below the `condensate´ propagators) are given by Dirac's
delta functions of the momenta. This circumstance produces
singular Feynman diagrams in the perturbation series even after
dimensional regularization is introduced. The singularities
correspond with the appearance of delta functions or standard
Feynman propagators evaluated at zero momentum, after some loop
integrals are performed. These factors occurs because of  the
momentum conservation in the vertices of the diagrams. Let us
consider a vertex having $n$ legs ($n=3,4$ for QCD). Then, when
$n-1$ different condensate lines join to it, the momentum
conservation forces the value of the momentum at the only
remaining line to vanish. Therefore, if a condensate line is
attached to this ending, delta functions evaluated at zero
momentum will appear. On the other hand, when a usual Feynman
propagator is connected to this last leg, a factor equal to its
value at zero momentum appears. This situation should be solved
before a sense could be given to the modified expansion. Below, we
propose two ways for the elimination of these singularities, which
clearly should be the subject of further examination about its
consistency. Let us separately consider in what follows the two
cases. It will be assumed that the Feynman diagrams are
constructed in $D$ dimensions and for a fixed value of the Feynman
parameter $\epsilon.$ Before advancing, let us precisely set the
rules for the order of the limits associated to the various
regularizations that have been done up to now:

\textit{a) The large volume limit will be taken by taking the size of the
quantization box }$L$\textit{\ going to infinity. This limit will lead to
continuous values of the momenta.}

\textit{b) The Nakanishi infrared regularization parameter }$\sigma $\textit{%
\ will be chosen as a function }$\sigma(L)$\textit{\ of the
spatial size of the system }$L,$\textit{\  vanishing as
}$L\rightarrow\infty .$\textit{\ This constant sets to zero all
the Feynman propagators for the spatial momenta lying within a
sphere of radius terms }$\sigma,$\textit{\ and all values of the
temporal component }$k_0$.

\textit{c) The maximal size }$P$\textit{\ of the set of the spatial momenta }%
$\vec{k}_{i}$\textit{\ for which condensate states are created will be
assumed to be another function }$P(L)$\textit{\ also vanishing in the limit }%
$L\rightarrow\infty.$\textit{\ However, it will be chosen smaller than }$%
\sigma(L),$\textit{\ in a proportion to be defined in the
following.}

\textit{d) Finally, the auxiliary mass }$m$\textit{\ of the quark
field will be chosen as a function }$m(L)$\textit{\ tending to
zero for }$L\rightarrow \infty$\textit{\ with a behavior to be
also described  below. }

The limit $L\rightarrow\infty,$ in which $\sigma(L)$, $P(L)$, and
$m(L)$ vanish, will be taken in the first place. With this step we
recover the Lorentz invariance of the Feynman diagrams that
continue to be functions of the dimensional regularization
parameter $D$, it mass scale $\mu$, and the Feynman constant
$\epsilon$.

\subsection{$\delta(0)$\ $\ $singularities}

An idea that comes directly to the mind when considering the
singular terms of the form $\delta (0)$ is that they should be
considered in dimensional regularization. \ \ But, it has been
argued that the delta-functions evaluated a zero spatial
coordinates can be analytically extended to continuous dimensions
$D$ and, moreover, that their expressions vanish in the limit $\
D\rightarrow 4$ \cite{liebbrandt}. This is not \ necessarily an
un-natural result, as it will \ illustrated in a second
method to remove these singularities, already in the step when the limit $%
L\rightarrow 0$ \ is taken. \ Helpful in this purpose will be a proper
selection of the set of momenta $\{k_i\}$ for which the condensate states
are created. \ Let us start, however, with the dimensional regularization
argument. \ We follow the same procedure here and interpret the Delta
functions appearing as $D$-dimensional ones. The Wick rotation of the
momentum integrals will be assumed to be already done. Then, it is possible
to reproduce, step by step, the arguments of Capper and Leibbrandt \cite
{liebbrandt} to conclude that these factors should vanish after removing the
dimensional regularization. Let us do it below for the sake of concreteness.
For $\delta (0)$ we can write
\[
\delta (0)=\int_E\frac{dp^D}{(2\pi )^D}.
\]
This is a singular $D$-dimensional integral in Euclidean momentum space,
similar to those in real space, considered in \cite{liebbrandt}. Then, it
can also be written as follows
\[
\int_{E}\frac{dp^{D}}{(2\pi)^{D}}=\int_{E}\frac{dp^{D}}{(2\pi)^{D}}\frac
{p^{2}}{p^{2}}=\int_{0}^{\infty}ds\int_{E}\frac{dp^{D}\,p^{2}}{(2\pi)^{D}}
\exp(-s\,p^{2}).
\]
However, using the redefinition of the generalized Gaussian integral for
continuous values of the dimension $D$ constructed in Ref. \cite{liebbrandt}%
, it is possible to write
\begin{eqnarray*}
\int_{E}\frac{dp^{D}\ }{(2\pi )^{D}}\exp (-s\,p^{2})
&=&\frac{1}{(4\pi )^{
\frac{D}{2}}}\exp \left[ -s\,f\left( \frac{D}{2}\right) \right] , \\
\int_{E}\frac{dp^{D}\ p^{2}}{(2\pi )^{D}}\exp (-s\,p^{2})
&=&\frac{1}{(4\pi )^{\frac{D}{2}}}\left[
\frac{D}{2}s^{-(1+\frac{D}{2})}+s^{-\frac{D}{2}}\ f\left(
\frac{D}{2}\right) \right]\exp \left[ -s\,f\left(
\frac{D}{2}\right) \right] ,
\end{eqnarray*}
and
\[
\delta (0)=\int_{0}^{\infty }ds\frac{1}{(4\pi )^{\frac{D}{2}}}\left[ \frac{D%
}{2}s^{-(1+\frac{D}{2})}+s^{-\frac{D}{2}}\ f\left(
\frac{D}{2}\right) \right] \exp \left[ -s\,f\left(
\frac{D}{2}\right) \right]
\]
where $f$ \ is the function introduced in \cite{liebbrandt} for extending
the generalized Gaussian integral formula to non-integral dimension
arguments. These functions vanish for all integral values of $D.$ Then,
after using the integral definition of the Gamma function $\ \Gamma
(z)=\int_0^\infty dt\ t^{z-1}\exp (-t),$ it follows that
\[
\delta(0)=\frac{f(\frac{D}{2})^{D}}{(4\pi)^{\frac{D}{2}}}\left[
\frac{D} {2}\Gamma\left( -\frac{D}{2}\right) +\Gamma\left(
1-\frac{D}{2}\right) \right],
\]
which vanishes exactly in the limit $D\rightarrow 4.$ Therefore, we will
assume that the evaluations associated to the modified expansion will be
made by using the above representation for the factors $\delta (0).$ As a
consequence, it will be considered that all the diagrams in which such a
kind of singularities appear will vanish in dimensional regularization.

Let us consider now a property through which it seems that we can
directly eliminate such singularities, even in the absence of the
above exposed dimensional regularization argument. The main
observation is that all the 4-momenta of the modes associated to
condensate states are null 4-vectors
lying in the forward light cone. Therefore, if we define the set $\{%
\overrightarrow{k}_i\}$ of the spatial momenta components used in
the construction of the `squeezed´ states,  with the unique
condition that all of them have different orientations in their
spatial parts, the sum of any subset of three or four \ 4-momenta
can not add up to a zero result, since the temporal components are
by construction all positive in the forward light cone. Therefore,
the imposition of the rigorous conservation of the 4-momentum can
not be satisfied by the sum of any set of three or four momenta
arriving to a given three or four legs vertex. Hence, this
observation indicates that the elimination of the $\delta (0)$
singularities is in fact a natural result of the way in which this
massless  problem has been regularized, which leads to the same
outcome given  by  dimensional regularization.

\subsection{b)\ Feynman propagator at\ $p=0\ \ $singularities}

For the elimination of this kind of singular behaviour, the Nakanishi
infrared regularization implemented on the Feynman propagators in a previous
section is specially helpful. These singular terms are simply vanishing
because of the appearance of the infrared regularizing factor \ $\theta (|%
\overrightarrow{p}|-\sigma )$ in each of the Feynman propagators,
before taking the limit $L\rightarrow \infty.$

Let us examine below the loop expansion after assuming the above
prescriptions and the modified free propagators for a gauge parameter $%
\alpha :$
\begin{align}
D_{g\mu \nu }^{ab}(p,m)& =\frac{\theta (|\overrightarrow{p}|-\sigma )\delta
^{ab}}{p^2+i\epsilon }\left( g_{\mu \nu }-\frac{(1-\alpha )p_\mu p_\nu }{%
p^2+i\epsilon }\right) -iC_g\ \delta ^{ab}\delta (p),  \label{free} \\
G_q^{^{f_1f_2}}(p,M,S)& =-\frac{\theta (|\overrightarrow{p}|-\sigma )\delta
^{f_1f_2}p_\mu \gamma ^\mu }{p^2+i\epsilon \ }-i\text{ }\delta ^{f_1f_2}C_q\
\delta (p),  \nonumber \\
G_{gh}^{ab}(p)& =-\frac{i\theta (|\overrightarrow{p}|-\sigma )\delta ^{ab}}{%
p^2+i\epsilon },  \nonumber
\end{align}
and the standard vertices of QCD. Then, it follows that all the diagrams
having a fixed number of loops showing both types of singularities will
vanish in the limit $L\rightarrow \infty .$ \ Therefore, after taking this
limit, the remaining finite diagrams (due to the dimensional regularization)
can be evaluated by using non-distorted dimensionally regularized
propagators, following from (\ref{free}) in the limit $\sigma \rightarrow 0$.

Assuming that the chosen conditions have eliminated all the singular
contributions, it is possible to comment about some properties of the
diagram series based on the above propagators. For example, it follows that
the appearance of a number $m$ of the condensate propagators within an $n$%
-loop one-particle-irreducible(1PI) diagram will eliminate $m$ of the $n$%
-loop integrals associated to this contribution. Therefore, the considered
diagram will now be an `effective'\ $(n-m)$-loop one. Consider also,
expressing the condensate parameters $C_g$ and $C_q$ in favour of the new
ones:
\begin{equation}
m^2=-\frac{6g^2C_g}{(2\pi )^4},\,\,\,\,\,\,S_f=\frac{g^2C_q}{4\pi ^4}\ ,
\label{newpar}
\end{equation}
which incorporate a power of 2 factor of the coupling constant $g.$ \ Then,
the $n$-loop 1PI diagrams of the effective expansion, considered as power
series in the new three parameters $m^2,$ $S_f$ and $g$, also show the
property that, given the number of external legs of the diagram, the number
of loops is fixed by the power of $g^2$ appearing in it. This conclusion
directly follows from the fact that each time a condensate internal line
appears, a loop integral is annihilated, and correspondingly the power of $g$
of the diagram is reduced by 2 in the new expansion. Let us consider that $%
P_g$ is the power of $g$ corresponding to an $n-$loop 1PI diagram in which
the parameters are the original ones. Thus the new power of $g$ of this
diagram with $m$ condensate lines when the new parameters are introduced
will be
\begin{equation}
P_g^{\prime }=P_g-2m.  \label{power}
\end{equation}
It seems that this property, can allow for a useful reordering of the
perturbation expansion. To see an indication for this, let us consider a
particular $n$-loop 1PI diagram in which the change of the parameters (\ref
{newpar}) have been introduced and corresponding line symbols have been
assigned for the standard and the condensate propagators separately. Then,
for any particular standard type line in this diagram, consider the infinite
summation of all the zero order in $g$ (tree) contributions to the connected
propagator. Any of the added terms, by construction have the same number of
loops, but comes from higher loops in the original expansion. This is done
by considering fixed the other standard lines. Thus, if not stopped by any
difficulty associated to the combinatorial and symmetry factors in the
diagrams, this infinite addition seems can be done for all the normal lines.
The resulting modified (zero order in the new expansion) propagators are no
other things that the ones employed in Ref. \cite{jhep} in the gauge $\alpha
=1$. Therefore, it seems possible to demonstrate that the loop expansion can
be reordered to produce an alternative version with modified propagators.
The investigation of this possibility will considered in next works.

\section{WTS identities}

\ \ \ This section is devoted to showing that the \ generating functional
constructed with the modified propagators, including condensate terms and
the gauge parameter \ $\alpha ,$ satisfies \ exactly the same WTS identities
as those associated to the usual PQCD. \ Let us consider for the cited
purpose \ the \ complete generating functional of the theory, \ which can be
written in the form
\begin{eqnarray}
Z\left[ j,\eta ,\bar{\eta},\xi ,\bar{\xi}\right]  &=&\exp \left\{ i\left(
\frac{S_{ikl}^g}{3!i^3}\frac{\delta ^3}{\delta j_i\delta j_k\delta j_l}+%
\frac{S_{iklm}^g}{4!i^4}\frac{\delta ^4}{\delta j_i\delta j_k\delta
j_l\delta j_m}+\right. \right.   \label{funct2} \\
&&\left. \left. \frac{S_{ikl}^q}{i^3}\frac{\delta ^3}{\delta \bar{\eta}%
_i\delta J_k\delta \left( -\eta _l\right) }+\frac{S_{ikl}^{gh}}{i^3}\frac
\delta {\delta \bar{\xi}_i\delta j_k\delta \left( -\xi _l\right) }\right)
\right\} Z_0\left[ J,\eta ,\bar{\eta},\xi ,\bar{\xi}\right] ,  \nonumber
\end{eqnarray}
in \ which the concrete form of the action $S$ assumed and its conventions
are defined in the Appendix A, and DeWitt compact notation for the
integration and summation over the corresponding spatial, colour or spinor
indices was employed. The specific set of indices associated to the Latin
letter $i,k,l$ is specified in a natural way by the particular type of field
to which it is associated as a subindex. For example in $j_i$ the meaning is
$i\equiv (x,\mu ,a).$ The free generating functional (\ref{zt}) can also be
expressed as a mean value in the initial state $|\Psi \rangle $ of the
evolution operator $U$ of the interaction Lagrangian of the external
sources, as follows:
\begin{align*}
Z_0\left[ J,\eta ,\bar{\eta},\xi ,\bar{\xi}\right] & =\lim_{T\rightarrow
\infty }\sum_{n,n^{\prime }}\langle \Psi |U(+T,-T|J,\eta ,\bar{\eta},\xi ,%
\bar{\xi})|\Psi \rangle , \\
& =\langle \Psi |T\left[ T\exp i\int d^4x\,\left( j\widehat{A}+\overline{\xi
}\text{ }\widehat{c}+\overline{\widehat{c}}\xi +\overline{\eta }\text{ }%
\widehat{\psi }+\widehat{\overline{\psi }}\eta \right) \right] |\Psi \rangle
, \\
& =Z_0[0,0,0,0]\exp \left[ i\frac{j_iD_g^{ik}j_k}2+i\bar{\eta}%
_iG_{gh}^{ik}\eta _k+i\bar{\xi}_iG_q^{ik}\xi _k\right] .
\end{align*}

Consider now two complete bases of eigenstates $\,|q^{\pm }\rangle
$ of the field operators \ taken at the fixed times $T$ and $-T$ .
Thus, for these states
\begin{align*}
\widehat{A}_\mu ^a\,(\pm T,\overrightarrow{x})\,|q^{\pm }\rangle & \equiv
A_{q^{\pm },\mu }^a\,(\overrightarrow{x})\,\,|q^{\pm }\rangle ,\text{ \ \ }
\\
\widehat{c}^a(\pm T,\overrightarrow{x})\,|q^{\pm }\rangle & \equiv c_{q^{\pm
}}^a\,(\overrightarrow{x})\,|q^{\pm }\rangle ,\text{ \ \ }\widehat{\overline{%
c}}^a(\pm T,\overrightarrow{x})\,|q^{\pm }\rangle \equiv \overline{c}%
_{q^{\pm }}^a\,(\overrightarrow{x})\,|q^{\pm }\rangle , \\
\widehat{\psi }^a(\pm T,\overrightarrow{x})\,|q^{\pm }\rangle & \equiv \psi
_{q^{\pm }}^a\,(\overrightarrow{x})\,\,|q^{\pm }\rangle ,\text{ \ \ }%
\widehat{\overline{\psi }}^a(\pm T,\overrightarrow{x})\,|q^{\pm }\rangle
\equiv \overline{\psi }_{q^{\pm }}^a\,(\overrightarrow{x})\,\,|q^{\pm
}\rangle .
\end{align*}
The explicit construction of the field operators for gauge theory in the
holomorphic representation is done in Ref. \cite{Faddeev}. This
representation is the most convenient for the construction of the basis
states $|q^{\pm }\rangle $ \cite{sakita}. Inserting the completeness
relation at the instants $-T$ and $T$ \ leads to \
\begin{eqnarray}
Z_0\left[ j,\eta ,\bar{\eta},\xi ,\bar{\xi}\right]  &=&\lim_{T\rightarrow
\infty }\sum_{q,q^{+}}\langle \Psi |q^{+}\rangle \langle
q^{+}|U(+T,-T|J,\eta ,\bar{\eta},\xi ,\bar{\xi})|q^{-}\rangle \langle
q^{-}|\Psi \rangle ,  \nonumber \\
&=&\lim_{T\rightarrow \infty }\sum_{q,q^{+}}\langle \Psi |q^{+}\rangle \frac
1{\mathcal{N}}\int \mathcal{D}\Phi \text{ }\exp {\Huge [}i\int_{-T}^Tdt\text{%
\ }\int d\overrightarrow{x}{\huge (}\frac 12A_\mu ^a\left( \partial ^2g^{\mu
\nu }-\left( 1-\frac 1\alpha \right) \partial ^\mu \partial ^\nu \right)
A_\mu ^a  \label{zo1} \\
&&-i\partial _\nu \overline{c}^a\partial ^\nu c+\overline{\psi }(i\gamma
^\nu \partial _\nu -m)\psi +jA+\overline{\xi }\text{ }c+\overline{c}\xi +%
\overline{\eta }\text{ }\psi +\overline{\psi }\eta {\huge )}{\Huge ]}\langle
q^{-}|\Psi \rangle ,  \nonumber
\end{eqnarray}
where $\mathcal{D}\Phi $ means the path integral differential $\mathcal{D}%
\{A,c,\overline{c},\psi ,\overline{\psi }\}$, and the usual representation
for the \ matrix element $\langle q^{+}|U(+T,-T|J,\eta ,\bar{\eta},\xi ,\bar{%
\xi})|q^{-}\rangle $ of the evolution operator between eigenstates
of the
fields \ has been used \cite{Faddeev}; $\mathcal{N}$ is the usual source$-$%
independent normalization constant fixing $Z_0[0]=1.$ \ The $(q^{-}$,$q^{+})$
dependence of the functional integral is contained in the boundary
conditions of the field integration variables, which should be fixed to the
eigenvalues $A_{q^{\pm },\mu }^a\,,\,c_{q^{\pm }}^a\,,\,\overline{c}_{q^{\pm
}}^a\,,\psi _{q^{\pm }}^a\,,\,\overline{\psi }_{q^{\pm }}^a\,$at the times $%
\pm T,$ \ respectively. \

By inserting (\ref{zo1}) in (\ref{funct2}), and acting with the interaction
part of the Action (expressed in terms of the functional derivatives) on $%
Z_0,$ it follows that
\begin{align}
Z& =\exp {\Huge [}i{\LARGE (}\frac{S_{ikl}^g}{3!i^3}\frac{\delta ^3}{\delta
j_i\delta j_k\delta j_l}+\frac{S_{iklm}^g}{4!i^4}\frac{\delta ^4}{\delta
j_i\delta j_k\delta j_l\delta j_m}+\frac{S_{ikl}^q}{i^3}\frac{\delta ^3}{%
\delta \bar{\eta}_i\delta j_k\delta \left( -\eta _l\right) }+\frac{%
S_{ikl}^{gh}}{i^3}\frac \delta {\delta \bar{\xi}_i\delta j_k\delta \left(
-\xi _l\right) }{\LARGE )}{\Huge ]}{\small \times }  \nonumber \\
& \text{ }Z_0\left[ j,\eta ,\bar{\eta},\xi ,\bar{\xi}\right] , \\
& =\lim_{T\rightarrow \infty }\sum_{q,q^{+}}\langle \Psi |q^{+}\rangle \int
\mathcal{D}\Phi {\small \times }\exp {\Huge [}i\int_{-T}^T\int dtd%
\overrightarrow{x}{\Huge (}{\Large -}\frac 14F_{\mu \nu }^aF^{\mu \nu
,a}-\frac 1{2\alpha }\partial ^\nu A_\nu ^a\partial ^\mu A_\mu ^a-i\partial
_\nu \overline{c}^aD^{ab,\nu }c^b+ \label{ztot}\\
& \text{ }+\overline{\psi }(i\gamma ^\nu D_\nu -m)\psi +j_\mu A^\mu +%
\overline{\xi }c+\overline{c}\xi +\overline{\eta }\psi +\overline{\psi }\eta
{\Huge )}\text{\ }{\Huge ]}\langle q^{-}|\Psi \rangle .  \nonumber
\end{align}

It can be noted that performing the derivatives over the sources of the
interaction Lagrangian inside the integrand of the Gaussian functional
integral is equivalent to adopting a perturbative definition of the
functional integral for the interacting theory, as discussed in Ref. \cite
{Zinn-Justin}.

\ That is, the classical action \ defining the complete generating
functional \ as a functional integral is exactly the same as for usual PQCD.
\ \ \ The changes in the diagrammatic expansion are thus only associated to
the boundary conditions at $-T$ and $+T$. It should be noted that \ for the
case of \ PQCD the \ wave functions $\ \ \Phi (q^{\pm })=\langle q^{\pm
}|0\rangle $ \ are specific Gaussian functions of the field components. It
can be shown that the \ $\langle q^{\pm }|\Psi \rangle $ are also Gaussian
wave functions, but defined by a different quadratic form as the argument of
the exponentials.

\ \ \ Therefore, it can be concluded that all \ Ward identities of the usual
PQCD should also be satisfied in the modified expansion. Since the action is
identical, it \ therefore has the same BRST invariance transformation: \
\begin{align*}
\delta A_\mu ^a(x)& =\delta \lambda D_\mu ^{ab}c^b(x), \\
\delta c^a(x)& =-\frac g2\delta \lambda f^{abc}c^b(x)c^c(x),\text{ \ \ }%
\delta \overline{c}^a(x)=-i\delta \lambda \partial _\mu A^{\mu ,a}(x), \\
\delta \psi (x)& =ig\delta \lambda c^a(x)T_a\psi (x),\text{ \ \ \ }\delta
\overline{\psi }(x)=-ig\delta \lambda c^a(x)\overline{\psi }(x)T_a,
\end{align*}
which leads, \ thanks to its Jacobian, equal to the unit, and
after applied as a  change of variables in (\ref{ztot}), leads to
the functional form of the WTS identities:
\begin{align*}
& \int dx\left( j^{\mu ,a}\left( x\right) D_\mu ^{ab}[\frac \delta {i\delta
j(x)}]\frac \delta {i\delta \bar{\xi}^b(x)}+\frac g2\overline{\xi }^a\left(
x\right) f^{abc}\frac \delta {i\delta \overline{\xi }^a(x)i\delta \overline{%
\xi }^a(x)}-i\xi ^a\left( x\right) \partial _\mu \frac \delta {i\delta j_\mu
^a(x)}+\right. \\
& \left. ig\frac \delta {i\delta (-\eta ^r(x))}\frac \delta {i\delta \bar{\xi%
}^a(x)}T_a^{rs}\eta ^s(x)-ig\overline{\eta }^s(x)\frac \delta {i\delta \bar{%
\xi}^a(x)}T_a^{sr}\frac \delta {i\delta \overline{\eta }^r(x)}\right)
Z\left[ j,\eta ,\bar{\eta},\xi ,\bar{\xi}\right] =0.
\end{align*}
The satisfaction of these relations,\ implies that the mean values for
physical quantities should be gauge-invariant, as was demonstrated in
general form in Refs. \cite{boulware,dewitt, thooft}.

\ \ \ It should be remarked \ that the proof of the gauge invariance of the $%
S$ matrix \ heavily relies on the existence of a mass shell \ defining the
asymptotic states for the particles. \ \ But, precisely, QCD is expected to
show the absence of these exact mass shells for its elementary fields: the
quarks and gluons. \ As, has already been evaluated in Ref. \cite{epjc},
these mass shells for quark and gluons are not following in the first
approximation in the condensate parameters and coupling constant. \

In ending this section, \ it can be noticed that the above outcome was
suggested by the fact that the new propagators also are inverse kernels of
the equations of motion of the free fields in massless QCD, as indicated by
the relations
\begin{eqnarray}
\left[ k^{2}g_{\mu\alpha}-\left(  1-\frac{1}{\alpha}\right)
k_{\mu}k_{\alpha }\right] \left\lbrace
\frac{1}{k^{2}+i\epsilon}\left[ g_{\nu}^{\alpha}-\frac{\left(
1-\alpha\right)  k^{\alpha}k_{\nu}}{(k^{2}+i\epsilon)}\right]
-ig_{\nu}^{\alpha}
C_{g}\text{ }\delta(k)\right\rbrace &  =g_{\mu\nu},\\
-\gamma_{\mu}k^{\mu}\left[
\frac{-\gamma_{\mu}k^{\mu}}{k^{2}+i\epsilon}+iC_{f}
I\delta(k)\right] =I,\qquad k^{2}\left(
\frac{1}{k^{2}+i\epsilon}\right) =I.&
\end{eqnarray}

\section{Gauge invariance in order $g^{2}$}

\ Let us \ now study the validity of the general
gauge$-$invariance properties that were obtained by means of the
formal path$-$integral procedures. \ \ The new expansion now has
the three parameters: the coupling constant $g$ and the two \ new
independent and dimensional ones \ associated to the gluon and
fermion condensates. \ \ Therefore, the invariance properties for
a given quantity should be obeyed in each order of a triple Taylor
expansion in those constants. \ The \ validity of the $\alpha $
independence will be examined for the expansion of some quantities
up to the second order in $g$ and any order in the other two
constants. \ \ In the first place the transversality of the gluon
self-energy, which is one of the basic WTS \ identities, will be
shown to be satisfied by the gluon self-energy in the
above-mentioned orders. Afterwards, in \ the same approximation,
the dimensionally regularized effective action \ will also be
shown to be gauge-parameter-independent. In what follows, the wavy
gluon and straight quark lines appearing in the Feynman diagrams
without any addition will mean the complete free propagators,
including the Feynman as well as the condensate components. \ The
same lines including a central empty circle
will indicate \ the Feynman part of the propagators (incorporating its $%
\alpha $ dependence); and the lines showing \ a central black dot
will represent the condensate parts. A wavy line including a
central transversal cutting segment will mean the usual gluon
propagator in the Feynman gauge, having a $g_{\mu \nu }$
Lorentz-tensor structure. \ Finally, a wavy line, but including a
central collinear segment, will represent the `longitudinal' part
of the gluon propagator, \ showing the Lorentz tensor structure $\
(1-\alpha )p_\mu $\ $p_\nu .$ \ This term includes all the \
$\alpha $ dependence of the free propagators. \ \

\subsection{Transversality of $\ \ \Pi_{\mu\nu}$ in order $g^{2}$}

\ Figure \ref{fig1} \ shows the diagrams contributing to the
polarization operator up to the second order in $g.$ \ Therefore,
collecting the terms in these expressions having certain definite
order in each of the two condensate parameters will define the
expansion coefficients of the \ polarization tensor in the series
of the three parameters.\ \ All the terms of non-vanishing order
in the condensate parameters are associated to the last four
diagrams in Fig. \ref{fig1}.

\begin{figure}[h]
\epsfig{figure=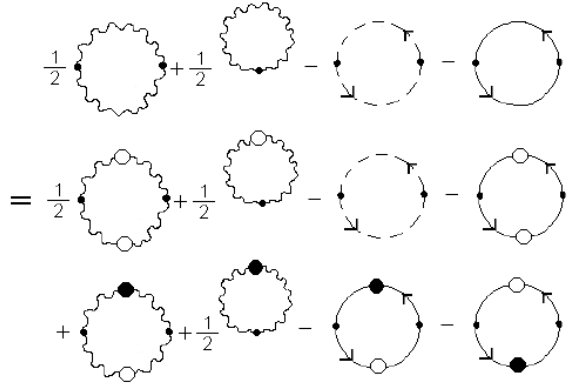,width=8cm} \vspace{-0cm} \caption{ The
diagrams contributing to the polarization tensor up to the second
order in $g$ and all orders in the condensate parameters}
\label{fig1}
\end{figure}
The first four diagrams in this figure \ represent the \ usual
second-order contribution to $\ \Pi _{\mu \nu }$ ,which is known
to be transversal. \ The condensate-dependent term \ can also be
represented as in Fig. \ref{fig2}.
\begin{figure}[h]
\epsfig{figure=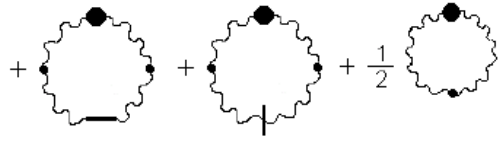,width=8cm} \vspace{-0cm} \caption{
Condensate dependent contributions to the polarization tensor. }
\label{fig2}
\end{figure}

Note the absence of terms coming form the quark condensate. This
is a consequence of the vanishing of these terms, which is
directly due to the fact that a trace of an odd number of $\gamma
$ matrices appears in their analytic expressions. \ \ \ \ Further
the last two diagrams in Fig. \ref {fig2} were the ones evaluated
in Ref. \cite{epjc} in the Feynman gauge and whose result is
transversal. \ Finally, the analytic expression of the first
diagram can be written in the form \

\[
T_{\mu \nu }=\frac{(1-\alpha )C_q}{(2\pi )^Di}%
(-g^2)f^{a_1a_2a}f^{a_1a_2b}V_{\alpha \beta \mu }(0,p,-p)g^{\alpha
\alpha
^{\prime }}V_{\alpha ^{\prime }\sigma \nu }(0,p,-p)\frac{p^\beta p^\sigma }{%
(p^2)^2},
\]
which, after employing the function defining the 3-gluon vertex
\cite{muta}
\[
V_{\mu _1\mu _2\mu _3}(k_1,k_2,k_3)=(k_1-k_2)_{\mu _3}g_{\mu _1\mu
_2}+(k_2-k_3)_{\mu _1}g_{\mu _3\mu _2}+(k_3-k_1)_{\mu _2}g_{\mu
_1\mu _3},
\]
leads to
\[
T_{\mu \nu }^{ab}=\frac{(1-\alpha )C_qg^2}{(2\pi )^Di}f^{a_1a_2a}f^{a_1a_2b}%
\left( g_{\mu \nu }-\frac{p_\mu p_\nu }{p^2}\right) ,
\]
showing the transversality of the polarization tensor up to second order in $%
g$ and all orders in the gluon and quark condensate parameters. \ \ \ The \ $%
\alpha $ \ independence of the effective action in the same
approximation will be studied below. \

\subsection{$\alpha -$independence of the effective action in order $g^2$}

\ \ \ Figure \ref{fig3} shows the \ terms of order 2 in the
coupling constant (and any order in the other two parameters) of
the effective action, evaluated at the zero value of the external
fields. Expanding the gluon propagator as a sum of its Feynman
gauge component plus the longitudinal and condensate parts; and
the quark one as the usual Feynman propagator plus the
quark$-$condensate part, the effective action can be expressed as
shown in Fig. \ref{fig3}.

\begin{figure}[h]
\epsfig{figure=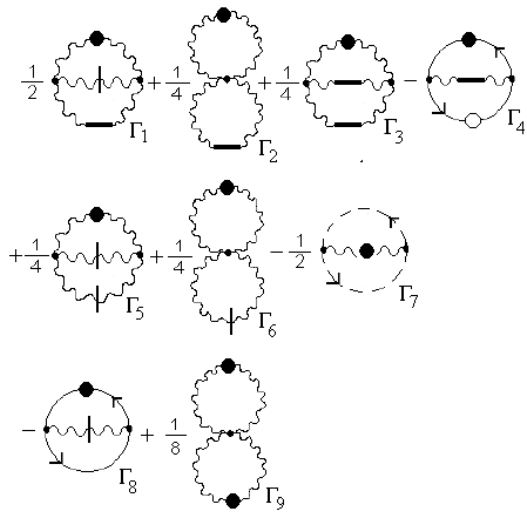,width=8cm} \vspace{-0cm} \caption{
Diagrams representing the contributions to the effective action in
the second order in $g$ and all order in the gluon and quark
condensates } \label{fig3}
\end{figure}

Observing this figure, it can be noted that the whole $\alpha
$-dependent contribution is given by the first four diagrams.
However, the first two of them vanish because the longitudinal
propagator is contracted with the polarization tensor in the
$g^{2\text{ }}$ approximation, which is transverse. Further, \ \
the third term also vanishes thanks to the fact that the
longitudinal propagator \ is contracted with the particular
self-energy part, which was shown, in the last subsection to be
transverse by itself. \ Finally, the \ self-energy quark term in
the fourth diagram identically vanishes, because of a trace of an
odd number of \ $\gamma $ \ matrices. This completes the proof of
the gauge invariance of the effective action in second order in
the coupling constant and any order of the condensate parameters.
\ \ \

In concluding this section it can be remarked that since massless
QCD has no initial dimensional parameter, the removal of the
dimensional regularization \ in the diagrams of the theory can be
done only after partial summations are performed. These
summations, then, will input the new dimensional parameters
associated to the condensates, implementing in this way the
dimensional transmutation effect \cite{colewein}. This and other
issues are expected to be addressed in next works.

\section{Summary}

The gauge-invariance and regularization properties of the perturbative
expansion for QCD considered in Refs. \cite{1995,PRD,epjc,jhep} was
investigated. If follows that the singularities produced by the Dirac delta-
function form of the new terms added to the gluon and quark propagators, can
be properly eliminated by combining dimensional regularization \cite
{liebbrandt} with the infrared regularization procedure for the operator
quantization of gauge theories \cite{Nakanishi}. The dimensional extension
allows to cancel the singular diagrams in which factors $\delta (0)$ appears
due to the momentum conservation laws at vertices, when all lines arriving
are associated to condensate $\delta $-like contributions. Further, the
Nakanishi infrared procedure allows to get rid of the singularities in the
form of a Feynman propagator evaluated at zero momentum.

Those factors appear when $n-1$ lines joined to a vertex of $n$
legs are of the condensate kind, and the remaining one is a
Feynman propagator. In connection with the gauge-invariance
properties, it is argued that the modified expansion should
satisfy the same WTS identities as the usual PQCD. In addition, it
follows that the formal functional integral representation of the
generating functional of the Green functions only differs from the
one associated to PQCD in the boundary conditions for the fields
at $t=\pm \infty $. The formal functional integral results are
checked in the second order in the coupling constant and any order
in the condensate parameters. Firstly, the transversality of the
gluon self-energy in the mentioned orders is shown. Afterwards,
the contributions to the effective action evaluated at vanishing
values of the mean fields, is shown to be
gauge-parameter-independent. The work is expected to be extended
in various directions. One of them is to implement the partial
summations of the diagrams, the possibility of what was advanced
in the text. This is needed in order to make the dimensional
transmutation effective, thus, allowing to perform calculation
depending on the new dimensional condensate parameters. Another
activity to be considered will be to specify the renormalization
procedure, which in the present problem involves three parameters.
Finally, a task of direct physical interest will be to incorporate
the knowledge about the invariance properties gained in the
present work. It could bring a better understanding the results
for the effective potential obtained in Ref. \cite{cabdann}, in
the Feynman gauge $\alpha =1$. The effective potential calculated
in this work gave signals of a strong instability of massless QCD,
under the generation of a quark condensate. But, it was precisely
the possibility of this outcome, that was the main motivation for
introducing the quark condensate in the initial free vacuum in
\cite {epjc,jhep}. Therefore, the check of the re-appearance of
the instability result within a gauge invariant calculation will
give further support to the dynamical generation of quark masses
and condensates within massless QCD.
\begin{acknowledgments}
One of the authors (A.C.) would like to deeply acknowledge the
invitation to visit the TH Division of CERN, in which this work
was completed. The helpful remarks and conversations there with
many colleagues as: F. Morales-Morera, S. Penaranda, R. Russo, M.
Luscher,  G. Altarelli, M. Mangano,  R. Fleischer, K. Rummukainen,
J. Bernabeu, R.K. Ellis, A. Martin,  G. Veneziano, P. Minkowski,
S. Pokorski, J. Ellis,  G. Corcella are also greatly appreciated.
Further,  the support of the ICTP for the travels and the
scientific activity related with the work, is also strongly
acknowledged. Finally, the friendly and efficient support of the
 Secretariat of the TH Division for one of the authors (A.C.) and
 the kind help of Suzy Vascotto in improving the
English of the manuscript are very much appreciated.
\end{acknowledgments}
\appendix

\section{Kugo-Ojima quantization procedure for QCD}

The main elements of the Kugo$-$Ojima operator quantization of the
free massless QCD for an arbitrary gauge parameter $\alpha $ will
be reviewed in this appendix. The formulae will be used in the
construction of the modified perturbative expansion done in
Section 2. \ The classical action for the interacting fields in
the Kugo-Ojima analysis has the form
\begin{eqnarray*}
S &=&\int dx\left( \mathcal{L}_g+\mathcal{L}_{gh}+\mathcal{L}_B+\mathcal{L}%
_q\right) , \\
\mathcal{L}_g &=&-\frac 14F_{\mu \nu }^a(x)F^{a\mu \nu }(x), \\
\mathcal{L}_{gh} &=&-i\partial ^\mu c^a(x)D_\mu ^{ab}(x)c^b(x), \\
\mathcal{L}_q &=&\overline{\psi }^r(x)(i\gamma ^\mu D_\mu ^{rs}(x)-m\delta
^{rs})\psi ^s(x), \\
F_{\mu \nu }^a(x) &=&\partial _\mu A_\nu ^a-\partial _\nu A_\mu
^a+gf^{abc}A_\mu ^bA_\nu ^c, \\
D_\mu ^{ab}(x) &=&\delta ^{ab}\partial _\mu -gf^{abc}A_\mu ^c, \\
D_\mu ^{rs}(x) &=&\delta ^{rs}\partial _\mu -gT_c^{rs}A_\mu ^c, \\
\lbrack T_a,T_b] &=&if^{abc}T_c.
\end{eqnarray*}

The equations of motion for the gluon, quarks, ghosts and auxiliary $B$
fields for a gauge parameter $\alpha $, in the Kugo and Ojima quantization
scheme for free massless QCD take the form \cite{Kugo,OjimaTex}
\begin{align}
{\ \partial^2 }A_\mu ^a\left( x\right) -\left( 1-\alpha \right) \partial_\mu
\text{ }B^a\left( x\right) & =0,  \nonumber \\
\partial ^\mu A_\mu ^a\left( x\right) +\alpha B^a\left( x\right) & =0,
\nonumber \\
{\ \partial^2 }B^a\left( x\right) ={\ \partial^2 }\text{ }c^a\left( x\right)
={\ \partial^2 }\text{ }\overline{c}^a\left( x\right) & =(i\text{ }\gamma
^\mu \partial _\mu \,-m)\psi (x)=0,
\end{align}
where, for the moment, quark fields are also considered as having an
auxiliary small mass $m$. The notation for the Lorentz indices and field
quantities will follow the ones used in Ref. \cite{muta}. \ In the case of a
general value of $\alpha $, the solution for the gluon field operator is the
only one that differs from its counterpart in the Feynman gauge $\alpha =1,$
which was considered in Ref. \cite{PRD}. \ As derived in \cite{Kugo,OjimaTex}%
, the non-vanishing commutation relations among the fields have the forms

\begin{align}
\left[ A_\mu ^a\left( x\right) ,A_\nu ^b\left( y\right) \right] &
=-\delta ^{ab}\left( g_{\mu \nu }D\left( x-y|0\right) -\left(
1-\alpha \right)
\partial _\mu \partial _\nu E\left( x-y\right) \right) ,  \nonumber \\
\left[ A_\mu ^a\left( x\right) ,B^b\left( y\right) \right] & =\delta
^{ab}\left( -\partial _\mu D\left( x-y|0\right) \right) ,  \nonumber \\
\left\{ c^a\left( x\right) ,\overline{c}^b\left( y\right) \right\} & =i\text{
}\delta ^{ab}D\left( x-y\text{ }|\,0\right) ,\text{ \ \ }\left\{ \psi \left(
x\right) ,\overline{\psi }\left( y\right) \right\} =\delta ^{rs}\text{ }%
(i\gamma ^\mu \partial _\mu +m)D\left( x-y\text{ }|\,0\right) ,  \nonumber \\
D\left( z\text{ }|\,m\right) & =\frac 1{(2\pi )^3}\int dk\text{
}s(k_0)\text{ }\delta (k^2-m^2)\exp (-ik.z),\text{ \ }E\left(
z\right) =\frac 12\frac 1{\nabla _z^2}\left( z_0\frac \partial
{\partial z_0}-1\right) \lim_{m\rightarrow 0}D\left( z\text{
}|\,m\right) ,  \nonumber  \\
  s(x)& =\theta(x)-\theta(-x). \nonumber
\end{align}

The invariant function $E$ appear in the commutation relations for the gluon
field \ when the \ gauge parameter differs from the Feynman gauge value $%
\alpha =1$, since not all the gluon fields satisfy the D'Alembert equation.
More generally, functions $E_{(\cdot )}$ are associated to each of the
invariant \ functions $D,D_{+},D_{-}$ and $D_F$ according to \
\begin{equation}
E_{(\cdot )}\left( z\right) =-\lim_{m\rightarrow 0}\frac \partial {\partial
m^2}D_{(\cdot )}\left( z\text{ }|\,m\right) \text{ }=\frac 12\frac 1{\nabla
_z^2}\left( z_0\frac \partial {\partial z_0}-1\right) \lim_{m\rightarrow
0}D_{(\cdot )}\left( z\text{ }|\,m\right) .  \label{E}
\end{equation}
where $(\cdot )$ means any \ of the subindexes of the functions \ $%
D,D_{+},D_{-}$ or $D_F.$

The gluon and Nakanishi $B$-field operators solving the above equations of
motion are given as \cite{Kugo,OjimaTex}\setlength\arraycolsep{0.5pt}
\begin{align}
A_\mu ^a\left( x\right) & =\sum\limits_{\vec{k}}\left( \sum\limits_{\sigma
=1,2}A_{\vec{k},\sigma }^af_{k,\mu }^{\text{ \ }\sigma }\left( x\right) +A_{%
\vec{k}}^{L,a}f_{k,L,\mu }\left( x\right) +B_{\vec{k}}^a\left[ f_{k,S,\mu
}\left( x\right) +(1-\alpha )\partial _\mu D_{(x)}^{(1/2)}g_k(x)\right]
\right) +\text{h.c.},  \label{sol1} \\
B^a(x)& =\sum\limits_{\vec{k}}B_{\vec{k}}^ag_k(x)+\text{h.c.},\text{ \ \ }%
c^a\left( x\right) =\sum\limits_{\vec{k}}c_{\vec{k}}^a\text{ }g_k(x)+h.c.,%
\text{ \ \ \ \ \ }\overline{c}^a\left( x\right) =\sum\limits_{\vec{k}}%
\overline{c}_{\vec{k}}^a\text{ }g_k(x)+\text{h.c.}, \\
\psi (x)& =\sum\limits_{\vec{k},s}\frac 1{\sqrt{2V}}(a_{\vec{k}}^s\text{ }u_{%
\vec{k}}^s\text{ }\exp (-i\text{ }kx)+b_{\vec{k}}^{s+}v_{\vec{k}}^s\text{ }%
\exp (+i\text{ }kx)), \\
\overline{\psi }(x)& =\sum\limits_{\vec{k},s}\frac 1{\sqrt{2V}}(b_{\vec{k}}^s%
\overline{v}_{\vec{k}}^s\text{ }\exp (-i\text{ }kx)+a_{\vec{k}}^{s+}\text{ }%
\overline{u}_{\vec{k}}^s\text{ }\exp (+i\text{ }kx)),\text{\ \ \ \ \ \ }
\end{align}
where h.c. means the Hermitian conjugate of the previous term. The wave
packets $g$ and $f$, the polarization vectors $\ \epsilon _\mu ^\sigma
\left( k\right) $, $\epsilon _{L,\text{\thinspace }\mu }\left( k\right) $, $%
\epsilon _{S,\,\mu }\left( k\right) ,$ the dirac spinor $u$ and $\ v$, and
the integro-differential operator \ $D_{(x)}^{(1/2)}$appearing in (\ref{sol1}%
) are defined as \cite{Kugo,OjimaTex}:
\begin{align}
g_k\left( x\right) & =\frac 1{\sqrt{2V\text{ }k_0}}\exp \left( -i\,kx\right)
,\quad \text{\ \ \ \ \ \ }f_{k,\mu }^\sigma \left( x\right) =\frac 1{\sqrt{2V%
\text{ }k_0}}\epsilon _\mu ^\sigma \left( k\right) \exp \left( -ikx\right)
\text{ \ \ },  \nonumber \\
kx& =k_0x_0-\vec{k}.\vec{x},\ \ k_0=\left| \vec{k}\right| ,  \label{paq} \\
\vec{k}\cdot \vec{\epsilon ^\sigma }\left( k\right) & =0,\ \epsilon
_0^\sigma \left( k\right) =0,\ \vec{\epsilon ^\sigma }\left( k\right) \cdot
\vec{\epsilon ^\tau }\left( k\right) =\delta ^{\sigma \tau },\text{ \ \ \
for \ }\sigma ,\tau =1,2, \\
\epsilon _{L,\mu }\left( k\right) & =-i\,k_\mu =-i\left( |\vec{k}|,-\vec{k}%
\right) ,\ \ \ \ \epsilon _{S,\mu }\left( k\right) =-i\frac{\overline{k}_\mu
}{2\,|\vec{k}|^2}=\frac{-i\left( |\vec{k}|,\text{ }\vec{k}\right) }{2\,|\vec{%
k}|^2},  \label{polar} \\
D_{(x)}^{(1/2)}& \equiv \frac 12(\nabla ^2)^{-1}(x_0\partial _0-1/2), \\
\,u_{\overrightarrow{k}}^s& \equiv \,u^s(k)=\sqrt{m+k_0}\begin{pmatrix}
u^{s}(0)\\
\frac{\overrightarrow{\sigma}\cdot\overrightarrow{k}}{m+k_{0}}u^{s}(0)
\end{pmatrix} ,\,\text{\ \ \ \ \ \ \ \ }v_{\overrightarrow{k}}^s\equiv
\,v^s(k)=\sqrt{m+k_0}\begin{pmatrix}
\frac{\overrightarrow{\sigma}\cdot\overrightarrow{k}}{m+k_{0}}v^{s}(0)\\
v^{s}(0) \end{pmatrix} ^T
\end{align}
The operator $D_{(x)}^{(1/2)}$ works as an ``inverse''\ of the D'Alembertian
for simple pole functions \cite{Nakanishi}, that is

\[
\partial^2 D_{(x)}^{(1/2)}f\left( x\right) =f\left( x\right) \text{ \quad if
\ \ }\partial^2 \,f\left( x\right) =0.
\]

A large cubic box of volume $V=L^3$ in a particular Lorentz reference frame
is assumed for the imposition of periodic boundary conditions on the fields.
Accordingly, the spatial momenta in the above sums take the values $\ \vec{k}%
=\frac{2\pi }L(n_1,n_2,n_3),$ for arbitrary integers $n_1,n_2$ $\ $and $n_3.$
The four-vectors for all the particles but the quarks are null ones, $k=(|\,%
\vec{k}\,|,\vec{k}),$ and for the quarks $k=(\sqrt{|\vec{k}\,|^2+m^2},\,\vec{%
k}).$ The table below shows the commutation relations between the creation
and annihilation operators for the various fields

$\ \ \ \ \ \ \ \ \ \ \ \ \ \ \ \ \ \ $%

\begin{equation}
\begin{Bmatrix}
& A_{\vec{k}^{\prime},\lambda^{\prime}}^{a^{\prime}+} &
A_{\vec{k}^{\prime} }^{L,a^{\prime}+} &
B_{\vec{k}^{\prime}}^{a^{\prime}+} & c_{\vec{k}^{\prime}
}^{a^{\prime}+} & \overline{c}_{\vec{k}^{\prime}}^{a+} &
a_{\vec{k}^{\prime}
}^{r^{\prime}+} & b_{\vec{k}^{\prime}}^{r^{\prime}+}\\
A_{\vec{k},\lambda}^{a}\ \ &
\delta^{aa^{\prime}}\delta_{\vec{k}\vec{k}^{\prime}
}\delta_{\lambda\lambda^{\prime}} & 0 & 0 & 0 & 0 & 0 & 0\\
A_{\vec{k}}^{L,a} & 0 & 0 &
-\delta^{aa^{\prime}}\delta_{\vec{k}\vec
{k}^{\prime}} & 0 & 0 & 0 & 0\\
B_{\vec{k}}^{a} & 0 &
-\delta^{aa^{\prime}}\delta_{\vec{k}\vec{k}^{\prime}} &
0 & 0 & 0 & 0 & 0\\
c_{\vec{k}}^{a} & 0 & 0 & 0 & 0 &
i\,\delta^{aa^{\prime}}\delta_{\vec{k}
\vec{k}^{\prime}} & 0 & 0\\
\overline{c}_{\vec{k}}^{a} & 0 & 0 & 0 &
-i\,\delta^{aa^{\prime}}\delta
_{\vec{k}\vec{k}^{\prime}} & 0 & 0 & 0\\
a_{\vec{k}}^{r} & 0 & 0 & 0 & 0 & 0 &
\delta^{rr^{\prime}} \delta_{\vec{k}\vec{k}^{\prime}}& 0\\
b_{\vec{k}}^{r} & 0 & 0 & 0 & 0 & 0 & 0 &
\delta^{rr^{\prime}}\delta_{\vec{k}\vec{k}^{\prime}}
\end{Bmatrix}.
\label{commu}
\end{equation}

\bigskip

\section{Longitudinal and Scalar Modes contribution}

The contribution of longitudinal and scalar modes  is determined
by the expression
\begin{eqnarray}
\langle 0 &\mid &\exp \left[ C_3^{*}\left( \alpha ,P\right) B_{\vec{p}
_i}^aA_{\vec{p}_i}^{L,a}+D^{*}\left( \alpha ,P\right) B_{\vec{p}_i}^aB_{\vec{
p}_i}^a\right] \times  \nonumber \\
&&\times \exp \left\{ i\int dxJ^{\mu ,a}\left( x\right) \left( A_{\vec{p}
_i}^{L,a+}f_{p_i,L,\mu }^{*}\left( x\right) +B_{\vec{p}_i}^{a+}\left[
f_{p_i,S,\mu }^{*}\left( x\right) +(1-\alpha )\left(
A^{*}\left(p_0,x_0\right) f_{p_i,L,\mu }^{*}\left( x\right) +B_\mu
^{*}\left( p_i,x\right) \right) \right] \right) \right\}  \nonumber \\
&&\times \exp \left\{ i\int dxJ^{\mu ,a}\left( x\right) \left( A_{\vec{p}
_i}^{L,a}f_{p_i,L,\mu }\left( x\right) +B_{\vec{p}_i}^a\left[ f_{p_i,S,\mu
}\left( x\right) +(1-\alpha )\left( A\left(p_0,x_0\right) f_{p_i,L,\mu
}\left( x\right) +B_\mu \left( p_i,x\right) \right) \right] \right) \right\}
\nonumber \\
&&\times \exp \left[ C_3\left( \alpha ,P\right) B_{\vec{p}_i}^{a+}A_{\vec{p}
_i}^{L,a+}+D\left( \alpha ,P\right) B_{\vec{p}_i}^{a+}B_{\vec{p}
_i}^{a+}\right] \mid 0\rangle .  \label{A1}
\end{eqnarray}

As in Refs.\ \cite{PRD,tesis}, we first calculate
\begin{eqnarray}
&&\exp \left\{ i\int dxJ^{\mu ,a}\left( x\right) \left( A_{\vec{p}%
_i}^{L,a}f_{p_i,L,\mu }\left( x\right) +B_{\vec{p}_i}^a\left[ f_{p_i,S,\mu
}\left( x\right) +(1-\alpha )\left( A\left( p_0,x_0\right) f_{p_i,L,\mu
}\left( x\right) +B_\mu \left( p_i,x\right) \right) \right] \right) \right\}
\nonumber \\
&&\times \exp \left[ C_3\left( \alpha ,P\right) B_{\vec{p}_i}^{a+}A_{\vec{p}%
_i}^{L,a+}+D\left( \alpha ,P\right) B_{\vec{p}_i}^{a+}B_{\vec{p}%
_i}^{a+}\right] \mid 0\rangle ,  \nonumber \\
&=&\exp \left\{ C_3\left( \alpha ,P\right) \left( B_{\vec{p}_i}^{a+}-i\int
dxJ^{\mu ,a}\left( x\right) f_{p_i,L,\mu }\left( x\right) \right) \right.
\times   \nonumber \\
&&\qquad \quad \times \left( A_{\vec{p}_i}^{L,a+}-i\int dxJ^{\mu ,a}\left(
x\right) \left[ f_{p_i,S,\mu }\left( x\right) +(1-\alpha )\left( A\left(
p_0,x_0\right) f_{p_i,L,\mu }\left( x\right) +B_\mu \left( p_i,x\right)
\right) \right] \right)   \nonumber \\
&&\quad \ \ \left. +D\left( \alpha ,V\right) \left( B_{\vec{p}_i}^{a+}-i\int
dxJ^{\mu ,a}\left( x\right) f_{p_i,L,\mu }\left( x\right) \right) \left( B_{%
\vec{p}_i}^{a+}-i\int d3xJ^{\mu ,a}\left( x\right) f_{p_i,L,\mu }\left(
x\right) \right) \right\} \mid 0\rangle .  \label{A2}
\end{eqnarray}
A similar expression is obtained acting to the left in Eq.\
(\ref{A1}). Substituting the above results in (\ref{A1}), and
introducing the following notation,
\begin{eqnarray}
C^{*} &\equiv &C_3^{*}\left( \alpha,P\right),\ C\equiv C_3\left(
\alpha,P\right),\quad D^{*}\equiv D^{*}\left( \alpha,P\right), \ D\equiv
D\left( \alpha,P\right),  \nonumber \\
\hat{A}^{+} &\equiv &A_{\vec{p}_i}^{L,a+},\ \hat{A}\equiv A_{\vec{p}
_i}^{L,a},\quad \hat{B}^{+}\equiv B_{\vec{p}_i}^{a+},\ \hat{B} \equiv B_{%
\vec{p}_i}^a,  \nonumber \\
a_1 &\equiv &-i\int dxJ^{\mu,a}\left( x\right) \left[ f_{p_i,S,\mu
}^{*}\left( x\right) +(1-\alpha)\left( A^{*}\left(p_0,x_0\right)
f_{p_i,L,\mu }^{*}\left( x\right) +B_\mu ^{*}\left( p_i,x\right) \right)
\right],  \nonumber \\
a_2 &\equiv &-i\int dxJ^{\mu,a}\left( x\right) \left[ f_{p_i,S,\mu }\left(
x\right) +(1-\alpha)\left( A\left(p_0,x_0\right) f_{p_i,L,\mu }\left(
x\right) +B_\mu \left( p_i,x\right) \right) \right],  \nonumber \\
b_1 &\equiv &-i\int dxJ^{\mu,a}\left( x\right) f_{p_i,L,\mu }^{*}\left(
x\right),  \nonumber \\
b_2 &\equiv &-i\int dxJ^{\mu ,a}\left( x\right) f_{p_i,L,\mu }\left(
x\right),  \label{nota}
\end{eqnarray}
one obtains
\begin{eqnarray}
&&\langle 0\mid \exp \left\{ C^{*}\left(\hat{A}+a_1\right) \left( \hat{B}%
+b_1\right) +D^{*}\left( \hat{B}+b_1\right) \left( \hat{B}+b_1\right)
\right\}  \nonumber \\
&&\times \exp \left\{ C\left( \hat{B}^{+}+b_2\right) \left( \hat{A}%
^{+}+a_2\right) +D\left( \hat{B}^{+}+b_2\right) \left( \hat{B}%
^{+}+b_2\right) \right\} \mid 0\rangle.  \label{1.34}
\end{eqnarray}

Following exactly the same procedure previously described in
Refs.\ \cite {PRD,tesis}, a recurrence relation is obtained for
Eq.\ (\ref{1.34}).
\begin{eqnarray}
&&\exp \left\{ C^{*}a_1b_1+D^{*}b_1^2+\left[ C\left( C^{*}a_1-a_2\right)
+2CD^{*}b_1\right] \left( C^{*}b_1-b_2\right) \sum\limits_{m=0}^n\left[
\left| C\right| ^{2\left( 2m\right) }+\left| C\right| ^{2\left( 2m+1\right)
}\right] +\right.  \nonumber \\
&&\quad +\left( C^{*}b_1-b_2\right) ^2\left( D\sum_{m=0}^n\left[ \left(
2m+1\right) \left| C\right| ^{2\left( 2m\right) }+2\left( m+1\right) \left|
C\right| ^{2\left( 2m+1\right) }\right] +\right.  \nonumber \\
&&\quad \quad \left. \left. +C^2D^{*}\sum_{m=0}^n\left[ 2m\left| C\right|
^{2\left( 2m-1\right) }+\left( 2m+1\right) \left| C\right| ^{2\left(
2m\right) }\right] \right) \right\} \langle 0\mid \exp \left\{ C^{*}\hat{A}%
\hat{B}+D^{*}\hat{B}\hat{B}\right\} \times  \nonumber \\
&&\times \exp \left\{ C^{*n+1}C^{n+1}\left( C^{*}b_1-b_2\right) \hat{A}
+\left[ C^{*n+1}C^{n+1}\left( C^{*}a_1-a_2\right)
+2C^{*n+1}C^{n+1}D^{*}b_1+\right. \right.  \nonumber \\
&&\qquad +\left. \left. 2C^{*n}C^n\left( n+1\right) \left(
C^{*}D+CD^{*}\right) \left( C^{*}b_1-b_2\right) \right] \hat{B}\right\} \exp
\left\{ C\hat{B}^{+}\hat{A}^{+}+D\hat{B}^{+}\hat{B}^{+}\right\} \mid
0\rangle.  \label{A4}
\end{eqnarray}

We then assume that $|C|<1$ so that in the limit $n\rightarrow \infty$
\begin{eqnarray}
&&\lim_{n\rightarrow \infty }\left| C\right| ^{2n}=\lim_{n\rightarrow \infty
}n\left| C\right| ^{2n}=0,  \nonumber \\
&&\lim_{n\rightarrow \infty }\sum\limits_{m=0}^n\left[ \left| C\right|
^{2\left( 2m\right) }+\left| C\right| ^{2\left( 2m+1\right) }\right] =\frac
1{\left( 1-\left| C\right| ^2\right) },  \nonumber \\
&&\lim_{n\rightarrow \infty }\sum_{m=0}^n\left[ \left( 2m+1\right) \left|
C\right| ^{2\left( 2m\right) }+2\left( m+1\right) \left| C\right| ^{2\left(
2m+1\right) }\right] =\frac 1{\left( 1-\left| C\right| ^2\right) ^2},
\label{lim}
\end{eqnarray}
which allows to rewrite Eq.\ (\ref{A4}) as
\begin{eqnarray}
&&\exp \left\{ C^{*}a_1b_1+D^{*}b_1^2+\left[ C\left( C^{*}a_1-a_2\right)
+2CD^{*}b_1\right] \frac{\left( C^{*}b_1-b_2\right) }{\left( 1-\left|
C\right| ^2\right) }+\left( C^{*}b_1-b_2\right) ^2\frac{\left(
D+C^2D^{*}\right) }{\left( 1-\left| C\right| ^2\right) ^2}\right\}  \nonumber
\\
&&\langle 0\mid \exp \left\{ C^{*}\hat{A}\hat{B}+D^{*}\hat{B}\hat{B}\right\}
\exp \left\{ C\hat{B}^{+}\hat{A}^{+}+D\hat{B}^{+}\hat{B}^{+}\right\} \mid
0\rangle .  \label{sus}
\end{eqnarray}
Replacing the compact notation (\ref{nota}) in (\ref{sus}), and expanding
all functions of $\vec{p}_i$ in the vicinity of $\vec{p}_i=0$ (keeping in
mind that the sources are located in a space finite region), it's sufficient
to consider only the first term in all expansions. After that Eq. (\ref{sus}%
) takes the form (\ref{LonSca}).

\end{document}